\pdfoutput=1
\documentclass[fleqn,usenatbib]{mnras}

\usepackage{newtxtext,newtxmath}

\usepackage[T1]{fontenc}

\DeclareRobustCommand{\VAN}[3]{#2}
\let\VANthebibliography\thebibliography
\def\thebibliography{\DeclareRobustCommand{\VAN}[3]{##3}\VANthebibliography}

\usepackage{graphicx}	
\usepackage{amsmath}	

\newcommand{\lya}{Ly$\alpha$} 
\newcommand{\hi}{\ifmmode {\textrm{H\textsc{i}}} \else H\textsc{i} \fi}
\newcommand{\hii}{\ifmmode {\textrm{H\textsc{ii}}} \else H\textsc{ii} \fi}

\title[Lyman $\alpha$ profiles]{Simulating the diversity of shapes of the Lyman $\alpha$ line}

\author[J. Blaizot et al.]{Jérémy Blaizot$^{1}$\thanks{E-mail: jeremy.blaizot@univ-lyon1.fr}, 
Thibault Garel $^{2,1}$, 
Anne Verhamme $^2$, 
Harley Katz$^3$,
Taysun Kimm$^4$,
\newauthor Léo Michel-Dansac$^1$,
Peter D. Mitchell$^1$,
Joakim Rosdahl$^1$,
Maxime Trebitsch$^5$
\\
$^{1}$Centre de Recherche Astrophysique de Lyon UMR5574, Univ Lyon, Univ Lyon 1, ENS de Lyon, CNRS, F-69230 Saint-Genis-Laval, France \\
$^2$ Observatoire de Gen\`eve, Universit\'ee de Gen\`eve, 51 Ch. des Maillettes, CH-1290 Versoix, Switzerland \\
$^3$ Sub-department of Astrophysics, University of Oxford, Keble Road, Oxford OX1 3RH, UK \\
$^4$ Department of Astronomy, Yonsei University, 50 Yonsei-ro, Seodaemun-gu, Seoul 03722, Republic of Korea \\
$^5$ Kapteyn Astronomical Institute, University of Groningen, P.O Box 800, 9700 AV Groningen, The Netherlands \\
}

\date{Accepted XXX. Received YYY; in original form ZZZ}

\pubyear{2023}

\begin{document}
\label{firstpage}
\pagerange{\pageref{firstpage}--\pageref{lastpage}}
\maketitle

\begin{abstract}

The \lya{} line is a powerful probe of distant galaxies, which contains information about inflowing/outflowing gas through which \lya{} photons scatter. To develop our understanding of this probe, we post-process a zoom-in radiation-hydrodynamics simulation of a low-mass ($M_{\rm star}\sim10^9\ M_\odot$) galaxy to construct 22500 mock spectra in 300 directions from $z=3$ to 4. Remarkably, we show that one galaxy can reproduce the variety of a large sample of spectroscopically observed \lya{} line profiles. While most mock spectra exhibit double-peak profiles with a dominant red peak, their shapes cover a large parameter space in terms of peak velocities, peak separation and flux ratio. This diversity originates from radiative transfer effects at ISM and CGM scales, and depends on galaxy inclination and evolutionary phase. Red-dominated lines preferentially arise in face-on directions during post-starburst outflows and are bright. Conversely, accretion phases usually yield symmetric double peaks in the edge-on direction and are fainter. While resonant scattering effects at $< 0.2\times R_{\rm vir}$ are responsible for the broadening and velocity shift of the red peak, the extended CGM acts as a screen and impacts the observed peak separation. The ability of simulations to reproduce observed \lya{} profiles and link their properties with galaxy physical parameters offers new perspectives to use \lya{} to constrain the mechanisms that regulate galaxy formation and evolution. Notably, our study implies that deeper \lya{} surveys may unveil a new population of blue-dominated lines tracing inflowing gas.

\end{abstract}



%

\section{Introduction}
Over the past decade, \lya{} emission from galaxies has quickly become one of the most important observational probes of the high-redshift Universe \citep[see e.g. reviews by][]{Barnes2014,Stark2016,Ouchi2020}. There are four primary reasons for this. First, most \lya{} photons are emitted by the interstellar medium (ISM) of star-forming galaxies, through recombinations of protons and electrons after hydrogen is photo-ionised by energetic radiation from short-lived massive stars. This fluorescent mechanism easily channels $\sim$~$5-10\%$ of the bolometric luminosity of these galaxies into the \lya{} line, which makes it an extremely bright spectral feature, ideal for detecting very faint objects in the high-$z$ Universe where \lya{} can be observed from the ground \citep{Partridge1967}. Second, the \lya{} transition has a large cross section -- even traces of neutral hydrogen between distant objects and the Earth will scatter \lya{} photons off the line of sight. This allows one to constrain the ionisation state of the intergalactic medium (IGM) with spectra of distant quasars \citep{Gunn1965}, or by measuring how the visibility of the \lya{} emission line from star forming galaxies varies with redshift \citep{Haiman2005} or position \citep{Furlanetto2006}. Third, because the \lya{} line is resonant, and again because of its large cross section, \lya{} photons will scatter many times through the ISM and circum-galactic medium (CGM) before they may reach the IGM and the observer. This scattering process introduces a coupling between the observational properties of the \lya{} line and the flows of gas it has traversed, which may be used to infer the physical conditions in galaxies and their environment from the \lya{} line shape \citep{Verhamme2006}. Fourth is technological progress. With the advent of a number of instruments which have made the spectroscopic observation and characterisation of the \lya{} line seem easy at all redshifts. Of particular interest are the Cosmic Origins Spectrograph (COS) onboard the Hubble Space Telescope \citep[HST/COS,][]{Green2012}, which has allowed exquisite spectroscopy of the \lya{} line of local star-forming galaxies \citep[see][and references therein]{Runnholm2021}, and the Multi-Unit Spectroscopic Explorer (MUSE) at the Very Large Telescope \citep[VLT/MUSE,][]{Bacon2010}, which has increased by orders of magnitude the number of high-quality \lya{} spectra of low-mass, star forming galaxies at redshifts between 3 and 6 \citep[e.g.][]{Inami2017}. 

It is thus clear that the \lya{} line is an extremely powerful and versatile observational tool, and that its observation at all redshifts has become largely accessible. The constraining power of these observations is now mostly limited by our theoretical understanding of the processes that shape the line emerging from galaxies and their CGM. Any model which addresses this question needs to propose an accurate description of (1) the sources of \lya{} emission, (2) the medium through which this radiation will propagate before reaching the observer, and (3) the process of resonant radiative transfer (RT), from the sources and through this medium. This latter point has received a lot of attention and may now be considered as a problem solved, at least numerically, in particular thanks to a number of powerful public \lya{} RT codes \citep[e.g.][and references therein]{Michel-Dansac2020}. Such codes have allowed to obtain numerical solutions to the \lya{} RT problem in the context of idealised models which describe more or less sophisticated flows of gas -- typically expanding shells -- around point sources \citep[see e.g. the pioneering works of][]{Dijkstra2006,Verhamme2006}. These models have proven to be impressively successful at reproducing the diversity of \lya{} line shapes \citep[e.g. ][and references therein]{Gronke2017,Gurung-Lopez2022} or at predicting the statistical \lya{} properties of distant galaxies \citep{Garel2012,Orsi2012}, and they have thus become the foundations of our interpretative framework. Despite their success, however, it is never clear to what extent these models are a faithful representation of reality, and to what extent the necessary simplifications they introduce to describe both the sources and the diffusing medium capture the physics going on in galaxies and their CGM. Indeed, constraints from these models appear to be degenerate \citep{Gronke2016,Li2022} and it is uncertain how their parameters relate to the physical conditions in galaxies. Numerical simulations of galaxy formation are the tool of choice to go beyond idealised models and make progress on points (1) and (2) above. The pioneering works of \citep{Tasitsiomi2006a,Laursen2007} have shown the importance of high spatial resolution and on-the-fly radiation hydrodynamics (RHD) to compute the non-equilibrium ionisation state of hydrogen in the simulated galaxies, and thus the source term of \lya{} radiation. \citet{Laursen2009} has further demonstrated the importance of accounting for dust in the computation of \lya{} RT, and proposed a simple and robust way to do so in post-processing of galaxy formation simulations. A number of numerical studies have followed \citep[e.g.][]{Yajima2012,Verhamme2012,Behrens2014}, but even the most recent attempts, using high resolution state-of-the-art simulations, have not been able to produce realistic \lya{} line profiles \citep{Behrens2019,Smith2019,Smith2022a,Smith2022b}. In particular, the fact that most observed \lya{} lines seem to be dominated by a red peak is hard to obtain with simulations. It is difficult to understand whether this is due to simulations failing to produce large-scale outflows \citep[see e.g. the discussion in][]{Smith2022a}, or whether it is a more subtle problem rooted at ISM scales. The crux of the problem clearly lies in producing a consistent and accurate description of both the sources of \lya{} radiation and the flows of gas from ISM to CGM scales. This requires not only high enough resolution, but also an adequate treatment of the necessary physics (radiation hydrodynamics) and sub-grid models for star formation and feedback from massive stars which produce the expected effect at the scales at which they operate without corrupting the thermodynamical properties of the ISM resulting from RHD. 

\vskip 0.2cm
In the present paper, we use a zoom-in, high-resolution, cosmological, radiation-hydrodynamics simulation of a typical high-redshift \lya{} emitter (LAE) to address the following three main questions:
\begin{itemize}
\item Can simulated galaxies produce the diversity of \lya{} line shapes that are observed, and in particular very red profiles (e.g. P-Cygni profiles) ? 
\item How do the properties of the \lya{} line vary in time and direction for a given galaxy?
\item What drives the shape of the \lya{} line, and, in particular, how is it connected to the physical properties of the ISM and CGM? 
\end{itemize}

The paper is organised as follows. In Sec.~\ref{sec:simulation}, we describe our simulation and how we construct mock observations from it. In Sec.~\ref{sec:ManyShapes} we confront our predictions to observations and discuss the diversity of \lya{} line shapes that our simulation produces. In Sec.~\ref{sec:SightLineVariations} we discuss how the mock lines vary in direction and time, and relate their properties to the flows of gas in the CGM. We carry out a short discussion in Sec.~\ref{sec:discussion} and conclude in Sec.~\ref{sec:conclusions}.

\section{Numerical methods and data}\label{sec:simulation}
The results presented in this paper are derived from the radiation-hydrodynamics, cosmological, zoom-in simulation of the low-mass, high-redshift galaxy described by \citet{Mauerhofer2021}, and which we have here post-processed with RASCAS \citep{Michel-Dansac2020} in order to obtain mock \lya{} spectra. We summarise the key aspects of our numerical methods below. 

\subsection{Simulation}
The simulation we use here is drawn from the sample of zoom-in simulations introduced in \citet{Mitchell2018}, and was re-simulated with the numerical methods of the {\sc Sphinx} project\footnote{http://sphinx.univ-lyon1.fr} presented in \citet{Rosdahl2018} \citep[see also][]{Mitchell2020}. Our simulation is fully described in \citet{Mauerhofer2021} and we only outline the methods here, referring the interested reader to these papers for the technical details. 

Our initial conditions are generated with MUSIC \citep{Hahn2011} to describe a dark matter (DM) halo of mass $M_{h} \sim 5\times 10^{10} M_\odot $ at $z=3$, in a periodic box of co-moving volume $20^3h^{-3}${\it c}Mpc$^3$, assuming a $\Lambda$CDM cosmology \citep{Ade2014}. The zoom-in high-resolution region approximately covers a sphere of radius $\sim 150$~kpc around the halo at $z=3$, with dark matter particles of mass $m_{\rm DM}\approx 10^4 M_\odot$. We have checked that there is no low-resolution DM particle within $3\times R_{\rm vir}$ of the halo position at any time. The simulation is performed with the adaptive mesh refinement code RAMSES \citep{Teyssier2002}, and employs a pseudo-Lagrangian refinement strategy to reach a  spatial resolution $\Delta x\approx 14$ pc where the density is highest. On top of this classical density refinement, we also refine the mesh in order to resolve the local Jeans length with at least four cells everywhere (down to the minimum cell width of $14$ pc). 

We use RAMSES in its radiation-hydrodynamics (RHD) version \citep{Rosdahl2013,Rosdahl2015,Katz2017,Rosdahl2018} to compute the interaction of ionising radiation with hydrogen and helium in real time. We follow \citet{Rosdahl2018} to model the injection and propagation of ionising radiation from star particles as they form and evolve using the BPASS-v2.0 model \citep{Eldridge2008,Stanway2016}. Because we resolve only a few galaxies in the simulation volume, we also include a uniform UV background (UVB) in order to reproduce reionisation. We use the redshift-dependent UVB model from \citet{Faucher-Giguere2009}, and account for self-shielding of the gas to the UVB by damping exponentially its intensity at densities above $n_H=0.01\ cm^{-3}$. Along with the non-equilibrium heating and cooling terms due to H and He, which are accounted for by the RHD module, we also include cooling from metals in a cruder way. At temperatures $T > 10^4$ K, metal cooling is derived from tables computed with CLOUDY \citep[][version 6.02]{Ferland1998} assuming a UVB from \citet{Haardt1996}. At lower temperatures, we use the fine-structure cooling rates from \citet{Rosen1995} and allow gas to cool down to $T=15$~K.

Thanks to the high resolution and to cooling below $T=10^4$~K, the ISM of our simulated galaxy naturally fragments to develop a population of star-forming molecular clouds. We use the subgrid model for star formation (SF) presented by \citet{Kimm2017} \citep[see also][]{Trebitsch2017a}, which is well adapted and tested in this regime. This model takes into account local turbulence to estimate the stability of a gas element and trigger star formation. It uses the multi-freefall formalism from \citet{Federrath2012} to define the local efficiency for the conversion of gas into stars. For supernova feedback, we use the mechanical feedback described in \citet{Kimm2014} and \citet{Kimm2015}. This model checks whether the local conditions allow the simulation to resolve the adiabatic expansion phase of SN explosions and injects energy and momentum accordingly, so that the snowplow phase either develops naturally or is imposed with momentum injection. For both the SF and feedback models, we have adopted the same parameters as used in the {\sc Sphinx} simulations presented in \citet{Rosdahl2018}, which were calibrated to recover the stellar-mass to halo-mass relation at high redshifts.

\vskip 0.2cm

\begin{figure}
	\includegraphics[width=\columnwidth]{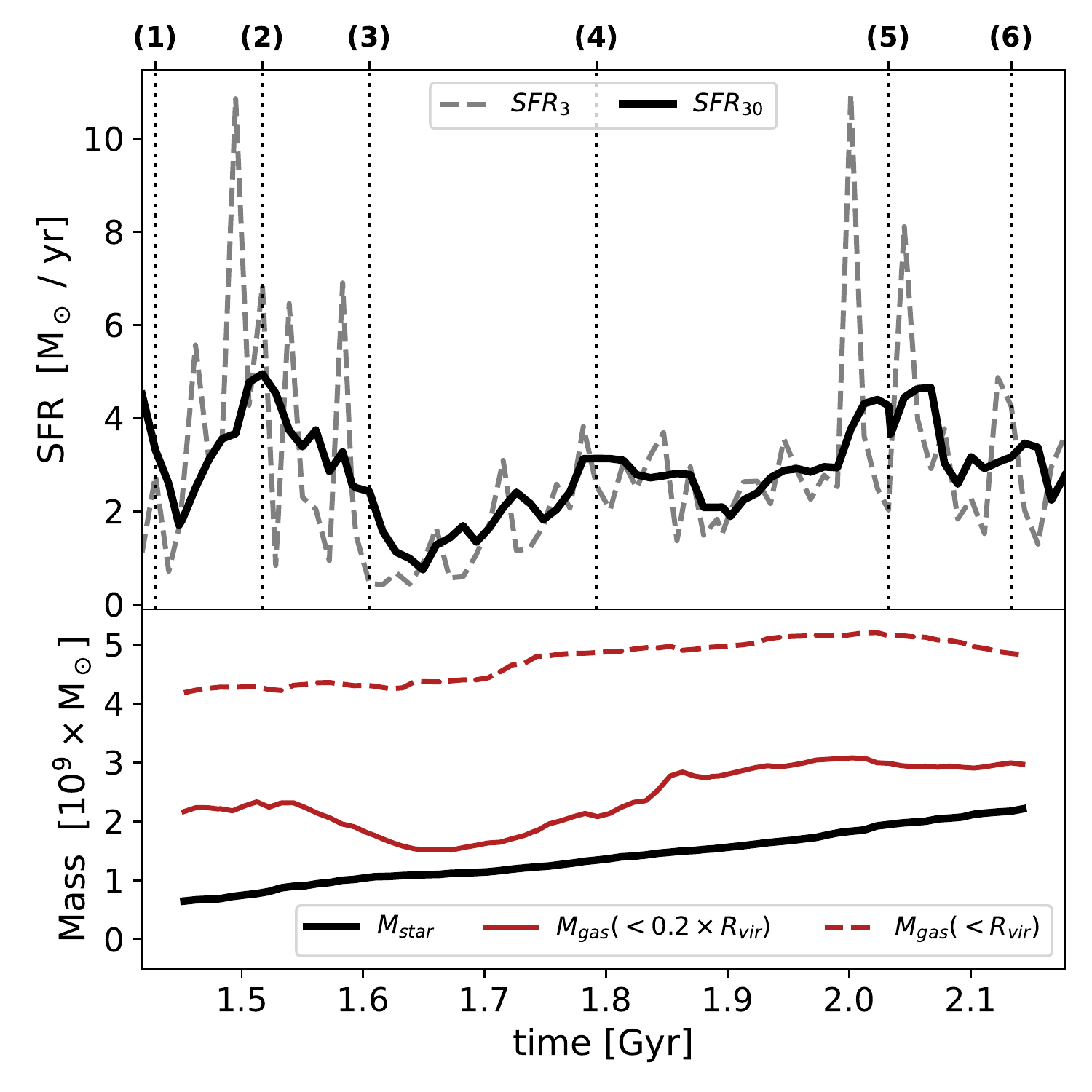}
    \caption{{\it Top panel:} Star formation rate history of our simulated galaxy. The solid curve shows the SFR averaged over 30 Myr, and the dashed curve shows SFR values counting stars formed over a 3 Myr window only. The vertical dotted lines mark specific times which are discussed in Sec. \ref{sec:SightLineVariations}. {\it Bottom panel:} The black line shows the stellar mass of the simulated galaxy as a function of time. The red lines show the masses of gas within $0.2\times R_{\rm vir}$ (solid curve) and within the virial radius (dashed line).}
    \label{fig:SFR_Mstar}
\end{figure}

We evolve the simulation down to redshift 3, and write outputs every 10~Myr. We identify and measure the properties of DM halos with the {\sc adaptahop} halo finder \citep{Aubert2004,Tweed2009} with the same parameters as used in \citet{Rosdahl2018}. In the present paper, we analyse 75 snapshots of the simulation taken between $z=4.18$ and $z=3$. In Fig. \ref{fig:SFR_Mstar}, we show the star formation rate and stellar mass histories of the galaxy during these $\approx 700$~Myr of evolution that we follow here. Over this period, the galaxy grows in stellar mass from $M_{\rm star} = 6.5\times 10^8\ M_\odot$ to $M_{\rm star} = 2.2\times 10^9\ M_\odot$, and its star formation rate, measured as the average over 30 Myrs, varies between $\approx 1$ and $\approx 5 M_\odot/{\rm yr}$ in a rather stochastic manner. At $z=3-4$, these values put the galaxy right on the star forming main sequence measured by \citet{Speagle2014}. The total mass of gas in the DM halo evolves smoothly from $4.2\times 10^9\ M_\odot$ to $4.9\times 10^9\ M_\odot$, with a maximum value of $5.2\times 10^9\ M_\odot$ at $t\approx 2$ Gyr. The mass of gas in the ISM and inner CGM (i.e. within $0.2\times R_{\rm vir}$) has stronger variations in time because of supernova feedback. In particular, roughly half the gas is ejected after the first star formation peak ($t\approx 1.6$ Gyr). The ISM mass then rebuilds until $t\approx 1.9$ Gyr, after which is it maintained roughly constant by the competition of accretion and feedback.

\subsection{Mock observations}

We use the code RASCAS \citep{Michel-Dansac2020} to construct mock \lya{} spectra from our simulated galaxy. We briefly discuss in Secs. \ref{sec:lyaEmission} and \ref{sec:Continuum} how we model the emission of \lya{} radiation and stellar continuum. In Sec. \ref{sec:gas_composition}, we describe how we model the gas and dust content through which radiation will propagate, and in Sec.~\ref{sec:rascas} we detail the mock observables that we construct. 

\subsubsection{\lya{} emission} \label{sec:lyaEmission}

\lya{} photons are mainly produced by radiative cascades that follow recombinations of H atoms. We compute the number of such \lya{} photons emitted per unit time from each cell of the simulation as
\begin{equation} \label{eq:LyaRecLum}
\dot{N}_{\rm rec} = n_e n_p \epsilon^B_{\rm Ly\alpha}(T) \alpha_B(T) \times (\Delta x)^3,
\end{equation}
where $n_e$ and $n_p$ are electron and proton number densities, $\alpha_{\rm B}(T)$ is the case B recombination coefficient, $\epsilon^B_{\rm Ly\alpha}(T)$ is the fraction of recombinations producing \lya{} photons, and $(\Delta x)^3$ is the volume of the cell. We evaluate $\epsilon^B_{\rm Ly\alpha}(T)$ with the fit from \citet[][their Eq. 2]{Cantalupo2008}, and $\alpha_{\rm B}(T)$ with the fit from \citet[][their appendix A]{Hui1997}.

\lya{} photons may also be produced through collisional excitations, and we evaluate this term with the recent computation from \citet[][their appendix A]{Katz2022} who include excitations up to level 5. These authors provide an accurate fitting formula to their results, which allows us to write the number of \lya{} photons emitted via collisional excitations per unit time from each cell as:
\begin{equation}\label{eq:LyaColLum}
\dot{N}_{\rm col} = n_e n_{\rm HI} \times \frac{(a / T^c) e^{-b / T^d}}{h\nu_\alpha}   \times (\Delta x)^3,
\end{equation}
where $n_{\rm HI}$ is the number density of neutral Hydrogen atoms, $h\nu_\alpha$ is the energy of the \lya{} photon in erg, and $(a,b,c,d) = (6.58\times 10^{-18},4.86\times 10^4, 0.185, 0.895)$ are the fit parameters from \citet{Katz2022}. We note that the emission rate from Eq. \ref{eq:LyaColLum} is very consistent with the similar estimate from \citet{Smith2022b}, and that both are significantly higher than the estimate from \citet{Goerdt2010}, by a factor $\sim 2$ at $T=10^5 K$, and less at lower temperatures. 

In both Eqs. \ref{eq:LyaRecLum} and \ref{eq:LyaColLum}, $n_{\rm HI}$, $n_p$, and $n_e$ are read directly from the output of the simulation which predicts the non-equilibrium ionisation state of H and He in each cell as a function of the local ionising radiation field. As can be seen from Eq. \ref{eq:LyaColLum}, \lya{} collisional emissivity is an extremely sensitive function of temperature. Predicting its value from numerical simulations thus demands a precise description of the thermal state of the gas \citep[see discussion in e.g.][]{Faucher-Giguere2010,Rosdahl2012}. It is unclear whether the thermal state of the gas in our simulation is accurately predicted in cells where the net cooling time is shorter than the hydrodynamical time-step set by the Courant condition. We thus use the conservative approach of \citet{Mitchell2020} and \citet{Garel2021} and set $\dot{N}_{\rm col}=0$ in cells where the net cooling time is shorter than five times the simulation timestep. As shown by \citet{Lee2022}, a slightly less conservative criterion already provides converged results in terms of H$\alpha$ emission. We check that the fraction of collisional emission which is resolved is typically more than 90\% at all times. This suggests that even if the un-resolved component were vastly underestimated (say by a factor 10), we would not make such a large error (a factor less than 2) on collisional emission, which remains sub-dominant relative to recombinative emission\footnote{We note that the cells where the cooling time is very short are generally cells with a low \lya{} escape fraction, as they are in very dense environments. The error on the observed \lya{} luminosity is thus in practice very low.}.

\subsubsection{Stellar continuum} \label{sec:Continuum}

Although the simulations were run using the version 2.0 of the BPASS model to compute the ionising emission of star particles, we use the updated version 2.2.1 of these models \citep{Stanway2018} to compute the non-ionising stellar continuum around the \lya{} line. In practice, the spectrum of each star particle is computed with a 2D interpolation in age and metallicity of the BPASS models, and then sampled to produce photon packets as described in Sec. 2.1.2 of \citet{Michel-Dansac2020}\footnote{Note that we use a slightly updated version of the RASCAS code, where the flux density $F_\lambda$ is now treated as a constant between BPASS data points instead of being linearly interpolated.}. 

In our experiments, we find that it is enough to compute the radiative transfer of stellar continuum within $\pm 5000$ km/s of the \lya{} line centre, and we thus emit continuum photons between 1170\AA{} and 1260\AA{} in the frame of each star particle. 

\subsubsection{H{\sc i}, deuterium, and dust} \label{sec:gas_composition}
The propagation of \lya{} photons from their emission site to the observer is determined by the neutral hydrogen, deuterium, and dust content on their path. The same holds for continuum photons around the \lya{} line, with a lesser impact of H{\sc i} (and D{\sc i}) further away from line centre. The number density of H{\sc i} in each simulation cell is a direct prediction of the simulation, and we use it as is. Following \citet{Dijkstra2006}, we model the number density of D{\sc i} ($n_{\rm DI}$) using a fixed ratio: $n_{\rm DI}/n_{\rm HI} = 3\times 10^{-5}$ \citep{Burles1998}. We model the dust content of each cell in post-processing, as described in  \citet{Michel-Dansac2020} \citep[see also][]{Mitchell2020,Mauerhofer2021}. Specifically, we implement the Small-Magellanic-Cloud (SMC) dust model of \citet{Laursen2009} and compute a pseudo-density of dust grains as $n_{\rm dust} = (n_{\rm HI} + 0.01 \times n_p) \times Z/0.005$, where $Z$ is the gas metallicity. We use the fits of \citet{Gnedin2008} for the SMC to compute the total cross section of dust, and use an albedo value of 0.32 \citep{Li2001}. 

The cross section we use for the \lya{} line is a Voigt profile with a width set by the thermal motion of H{\sc i} atoms and by the unresolved turbulent velocity distribution of the gas. We follow \citet{Mauerhofer2021} and use a uniform value of 20 km/s to describe this latter term.  

\subsubsection{RASCAS setup} \label{sec:rascas}

We construct 22,500 spectra by mock-observing the galaxy along 300 directions in each of the 75 simulation outputs. The directions are fixed and follow the HEALPix \citep{Gorski2005} decomposition with $n_{\rm side}=5$.  We use the peeling off technique \citep{Zheng2002,Dijkstra2017} to compute spectra in these 300 directions, collecting radiation only within an aperture of diameter 1 arcsec centred on the DM halo position. This aperture is always larger than a tenth of the virial radius (by 35\% at $z=3$ and 70\% at $z=4$), i.e. comparable to the size of the galaxy. The mock spectra have a fixed resolution in wavelength corresponding to a velocity resolution of $\delta v\sim 10$ km/s. 

We compute the emission and propagation of photon packets from and through a volume of radius $3\times R_{\rm vir} \sim 60 \ (90)$ kpc at $z=4.2\ (3)$. In the cosmology we assume, the Hubble flow at $3\times R_{\rm vir}$ is roughly constant from $z=4.2$ to $z=3$, and equal to $\sim 27$ km/s (or 9 km/s at the virial radius). This is comparable to the resolution of our mock spectra, and very small compared to features of the mock-observed \lya{} line. Contrary to \citet{Garel2021}, we thus choose to ignore the Hubble flow in our \lya{} RT computations with RASCAS for the present paper.

In practice, we carry out 3 independent RASCAS runs: one for collisional emission, where we use $2\times 10^5$ photon packets, one for recombinations, where we cast $5\times 10^5$ photon packets, and one for continuum photons, for which we use $10^6$ photon packets. Photon packets are emitted from cells (for \lya{} radiation) or star particles (for stellar continuum) with a probability proportional to the luminosities of the sources. The initial positions of \lya{} photon packets are drawn randomly assuming a uniform probability across each emitting cell. The initial positions of continuum photon packets are the positions of the emitting star particles. The frequencies of \lya{} photon packets sample a Gaussian distribution of width given by the local thermal velocity dispersion of the gas, and centred at the \lya{} frequency in the frame of the emitting cell. The directions of emission of all photon packets are drawn from an isotropic distribution. 

\vskip 0.2cm 
In order to define the systemic redshift of each mock observation and the width of the intrinsic \lya{} line, we also use RASCAS to compute the propagation of the \lya{} photon packets described above, assuming now that the medium is completely transparent. These cheap computations produce the intrinsic emission lines which contain the line-of-sight dependent kinematic imprint of the emission sites, but no scattering effect. We define the systemic velocity of the galaxies as the first moment of these spectra. 


\section{The many shapes of the \lya{} line}\label{sec:ManyShapes}
\begin{figure*}
	\includegraphics[width=\textwidth]{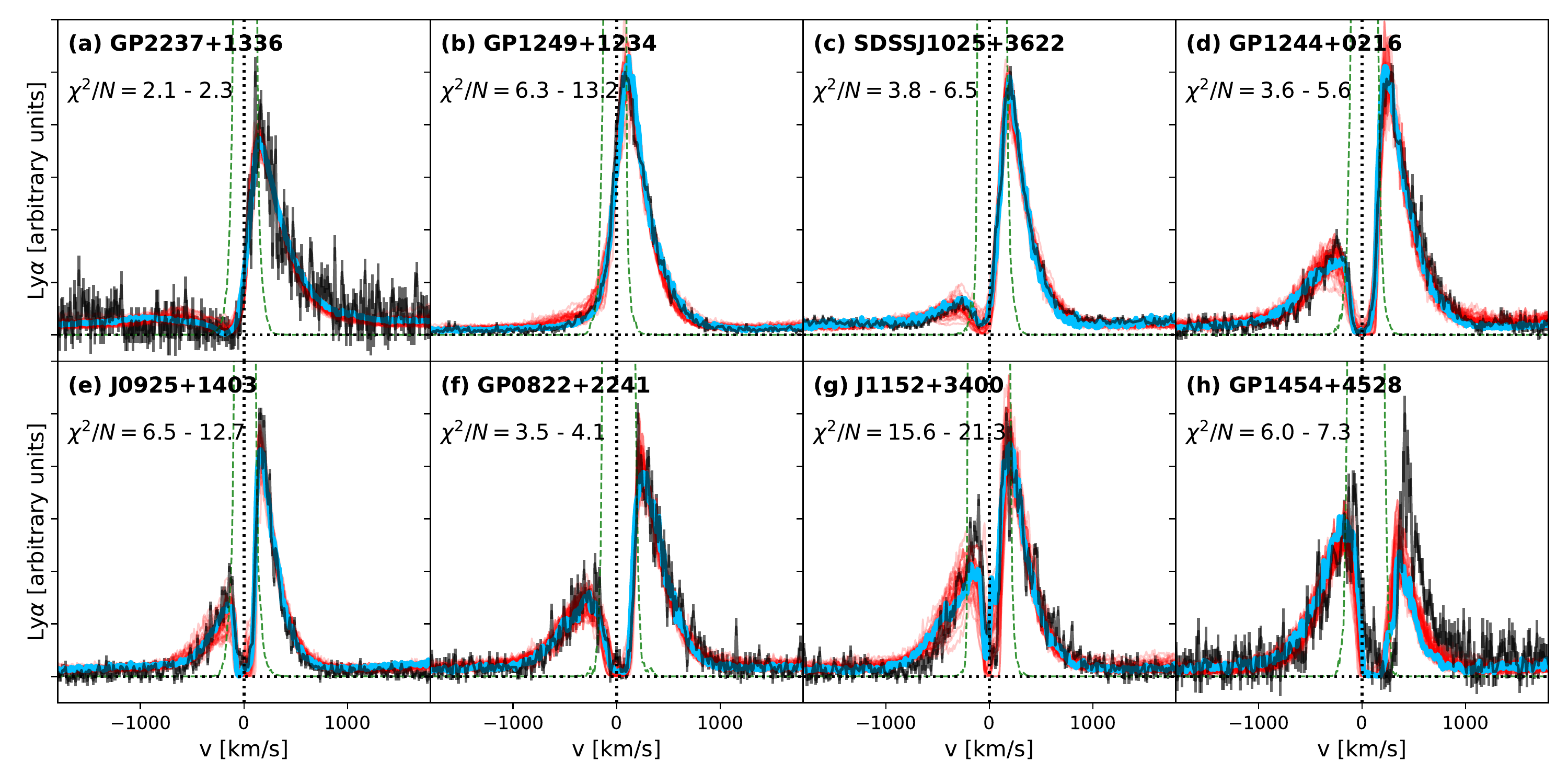}
    \caption{Selected examples of observed spectra from the LASD which are well reproduced by the mock spectra. In each panel, the observed spectrum is shown in black and the best-matching mock in blue. The bundle of thin red curves show the next 29 best matching spectra in our sample of mocks. The green dashed lines show the intrinsic emission corresponding to the best matched mocks. The name of the LASD galaxy is written on each panel for reference \citep[see][for the references]{Runnholm2021}. We write in each panel the values of the reduced $\chi^2$ corresponding to the best-match and 30th best-match mock spectra. }  
\label{fig:SuccessfulFits}
\end{figure*}

In this section we assess how our mock spectra reproduce the diversity of observed \lya{} line shapes. We start with a direct comparison of our mocks to observed \lya{} lines in Sec. \ref{sec:LASDComparison}. We then discuss more generally the distribution of line shapes that we find (Sec. \ref{sec:LyaShapes}) and how these mock lines populate parameter space (Sec. \ref{sec:LineParams}). 


\subsection{A comparison to observations}\label{sec:LASDComparison}

We compare our mocks to observations of low-redshift \lya{} lines taken from the {\it Lyman Alpha Spectral Database}\footnote{http://lasd.lyman-alpha.com/, as it was on April 25, 2022.} \citep[LASD,][]{Runnholm2021}. We focus on the low-redshift observations of the LASD because they generally have better signal-to-noise ratio and spectral resolution than high-redshift observations, and because they provide a robust measure of the continuum and hence of absorption features in the \lya{} line, which is often elusive in high-redshift LAE spectra. More importantly, these low-redshift lines are very unlikely to be affected by any intervening intergalactic gas, and thus inform us on the \lya{} line shape produced by galaxies. This makes them directly comparable to our mocks, which do not include any IGM transmission either. 

In practice, we selected from the LASD the 124 galaxies with redshifts lower than 0.5. The \lya{} spectra of these galaxies were obtained with COS onboard the Hubble Space Telescope and are described in \citet{Runnholm2021}. It is important to remember that this sample is large because it combines a number of independent surveys, but it is not designed to be complete or representative of the galaxy population in general.  In practice, the galaxies gathered in the LASD are a very heterogeneous sample, with star formation rates ranging from $0.1$ to $100\ M_\odot/yr$ and diverse selections on luminosity, morphology, colour, or spectral properties \citep[see Sec. 3.1 of][]{Runnholm2021}, and our simulated galaxy is indeed only comparable to a fraction of the LASD sample in terms of stellar mass and star formation rate. Nevertheless, this is a unique and very useful dataset to confront theoretical predictions, and certainly a successful model should be able to reproduce at least some of the line shapes compiled in the LASD. 

\vskip 0.3cm
We wish to find the mock spectra that are closest to each observed spectrum from the LASD. In order to do this, we introduce one single free parameter, which is the normalisation of each mock spectrum. We do not change anything else in the mocks other than this global normalisation. For example, the equivalent widths and systemic redshifts are fixed. We can then define model $M$ as $\alpha\times m_i$ where $\alpha$ is this free normalisation and $m_i$ are the values of the mock spectrum (interpolated to be at the same wavelengths as the observed ones, in the rest-frame). We compute the best value of $\alpha$ analytically for each pair of mock and observed spectra by minimising the $\chi^2$, i.e. requiring that $d\chi^2/d\alpha=0$. This yields $\alpha = (\sum_i m_i d_i /\sigma_i^2 )/ (\sum_i m_i^2/\sigma_i^2)$, where ${d_i}$ and ${\sigma_i}$ are the data points and associated errors from the observed spectrum, and the sums extend over a selected wavelength or velocity range. With this normalisation, we then compare each spectrum of the LASD to the 22,500 mocks by computing a single $\chi^2$ value for each pair of spectra. We define the best-matching mocks as those that produce the lowest $\chi^2$ values. In practice, the best matches will depend slightly on the velocity range over which we extend the comparison, as this gives more or less weight to continuum or line features. We have used [-1500; 1500] km/s as a default, and verified that our results are qualitatively unchanged when using slightly different ranges.

The reduced $\chi^2$ values we obtain are generally good, albeit somewhat high. Half of the best-matches yield values of $\chi^2$ below 5, and 20\% cases have reduced $\chi^2 > 15$. Upon visual inspection, however, we find that these scores are not a very satisfactory representation of what happens. Often, very low $\chi^2$ values tell us more about the noise level in the data than about the ability of the mock spectra to reproduce a particular line shape. Conversely, very large $\chi^2$ are sometimes obtained for matches that look very satisfactory by eye, and may be driven by surprisingly high signal-to-noise ratio in the observed spectrum, or features which are not associated to the \lya{} line and hence not modelled in our simulations (which compute the stellar continuum and the \lya{} transfer only).

\vskip 0.3cm
We show all the results of this matching procedure in Appendix \ref{app:allfits} and we focus here on selected examples. In Fig. \ref{fig:SuccessfulFits}, we show some of the spectra from the LASD that are well matched by our mocks. These examples were chosen to illustrate the diversity of line profiles in the LASD: they include a P-Cygni profile in panel (a), a single peak in panel (b), and double peaks with increasing flux in the blue in panels (c) to (h). The first striking impression from Fig. \ref{fig:SuccessfulFits} is that our mock spectra show a level of agreement with observations which is qualitatively at least as good as that obtained by fitting idealised models \citep[e.g.][]{Gurung-Lopez2022}. It is the first time to our knowledge that mock spectra from cosmological simulations reproduce so accurately observed \lya{} profiles. As shown with the bundle of thin red curves in each panel, the best-matching mock spectrum is generally not isolated, and the next 29 best matches provide a satisfactory representation of the observed spectrum as well. The reduced $\chi^2$ values are given on each panel and range between $\approx 2$ and $\approx 20$ for these fits. This is typical of the range of values we obtain for the full LASD sample, where more than 80\% best-matches are found with a reduced $\chi^2$ lower than 15. As can be seen in panels (b) and (g) of Fig. \ref{fig:SuccessfulFits}, such a high value is not necessarily worrying in terms of capturing the line shape properties faithfully. We note that the 30 mocks\footnote{We obtain similar results with 60 or 15.} that best represent each LASD spectrum are generally drawn from many snapshots of the simulation, and computed in different directions. We will understand from Sec. \ref{sec:SightLineVariations} that the link between the \lya{} line shape on the one hand, and the state of the galaxy and the direction of observation on the other hand, is indeed relatively tight only for extreme (and rare) line profiles, but generally rather loose.

\begin{figure}
	\includegraphics[width=\columnwidth]{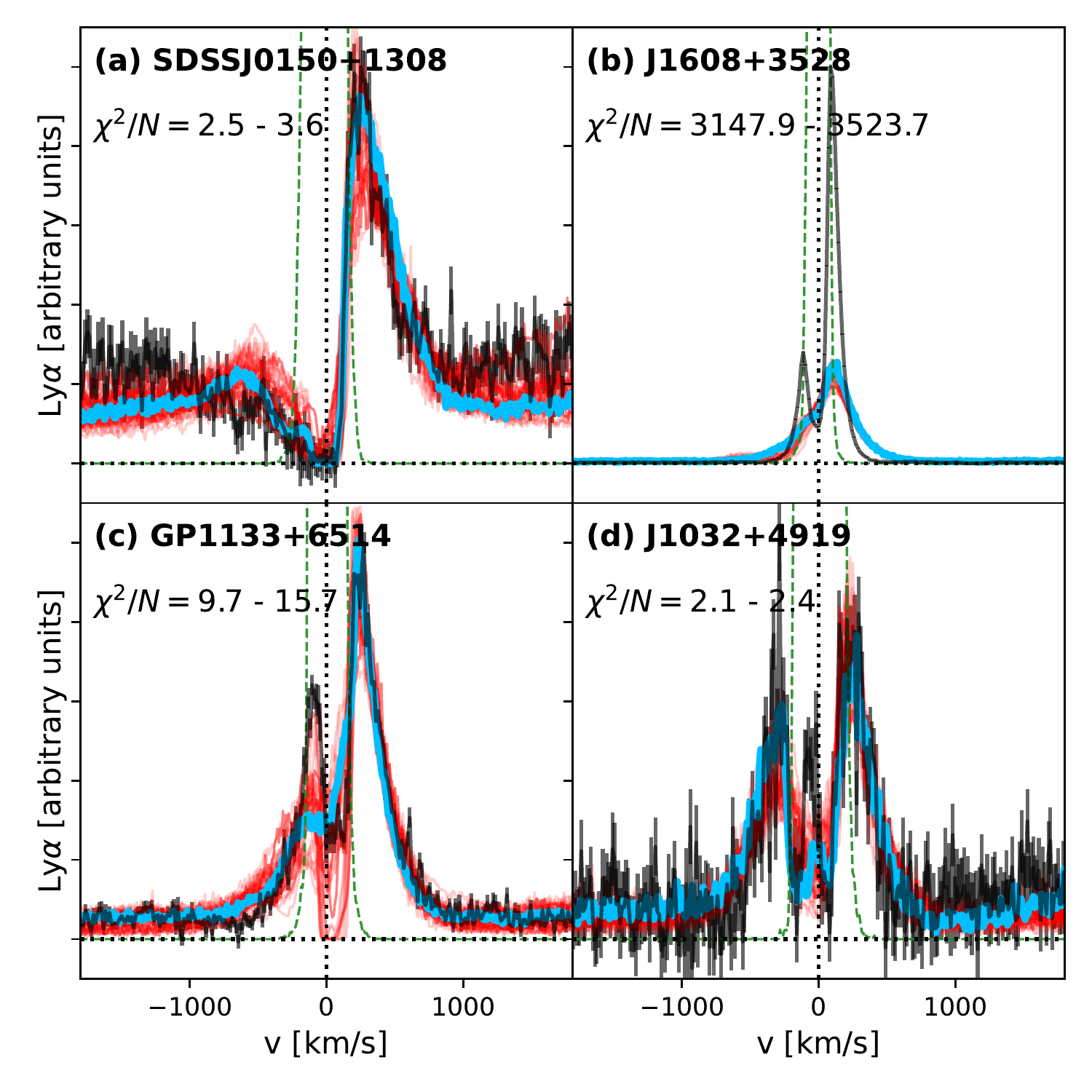}
    \caption{Same as Fig. \ref{fig:SuccessfulFits}, but with examples of observed spectra which have no satisfactory match in our sample of mock observations.}
    \label{fig:FailedFits}
\end{figure}

While a large majority of low-$z$ LASD spectra are well reproduced by our sample of mocks, it is interesting to discuss the minority which fails as well. In Fig. \ref{fig:FailedFits}, we show examples of observed spectra for which the simulation provides unsatisfactory matches. These 4 examples are chosen to highlight different types of problems. In {\it panel (a)}, we can see with the case of J0150+1308 \citep{Heckman2011} that despite an acceptable $\chi^2$ value, the matched mock spectra do not reproduce well the shape of the continuum away from the line. Indeed, there seems to be a general pattern where our mocks do not reproduce the broad absorption features which sometimes surround the \lya{} emission. Such features may in part come from the underlying stellar continuum of a relatively evolved stellar population \citep[see e.g.][]{Pena-Guerrero2013}, which would not be in place in our low-mass, high-redshift galaxy. More likely, though, the origin of the discrepancy is due to the gas content in the ISM and CGM of the galaxy. It is interesting that \citet{Heckman2011} report a stellar mass and star formation rate for J0150+1308 which are both about an order of magnitude larger than those of the galaxy we have simulated. We are thus comparing a Lyman-Break Analog (LBA) to a mock low-mass LAE, and it is not surprising that the LBA is found to have richer and faster outflows which produce broader absorption features \textcolor{blue}{\citep[see e.g.][]{Trainor2015}}.

In {\it panel (b)} of Fig. \ref{fig:FailedFits}, we show the example of the Green Pea galaxy J1608+3528 from \citet{Jaskot2017}, which has a very narrow \lya{} line with a high equivalent width and a rather shallow absorption feature. As underlined by these authors, this galaxy is an extreme in their sample, with by far the largest [O{\sc iii}]/[O{\sc ii}] ratio, very little signs of outflowing material, low column density of low-ionisation gas, and probably an extremely young population of massive stars in formation. It is perhaps not surprising that none of our mocks match this exceptional observation, which may correspond to a short-lived, hence rare, phase in the evolution of low-mass galaxies. Perhaps better time sampling in the outputs of the simulation would yield such rare events, but indeed, we find none in our sample. It is interesting that most of the \lya{} lines presented by \citet{Izotov2020} are also absent from our sample. These observations again target compact galaxies with extreme [O{\sc iii}]/[O{\sc ii}] ratios, most of which have a very narrow \lya{} line with high equivalent width, unmatched in our catalog. These narrow double peak profiles are associated to strong LyC leakage from observations \citep{Izotov2018,Flury2022}, and as shown in \citet{Mauerhofer2021}, our simulated galaxy is not a particularly strong LyC leaker, with an escape fraction well below 1\% most of the time and in most directions of observation. It may thus be that we miss the peculiar directions in which our galaxy would show narrow double peaks because these are more rare than 1 in 300. Or it may be that our galaxy is simply not an analog of the observed low-redshift Lyman continuum leakers which are admittedly rare. Cases like the one illustrated in panel (b) of Fig. \ref{fig:FailedFits} represent the majority of observed spectra which are not well matched by our mocks. It is interesting that these lines appear to be difficult to reproduce with simple shell models as well (see e.g. \citet{Orlitova2018} or \citet{Gurung-Lopez2022} whose best-fit models seem overestimate the absorption strength).

The cases of GP1133+6514 on {\it panel (c)} and J1032+4919 on {\it panel (d)} are different. These two \lya{} profiles show significant residual flux between the two peaks, which is in the form of a third emission peak close to (but not exactly at) systemic. It is clear that these two cases are peculiar sight-lines, which one cannot expect to fit in detail with our mocks. While there are indeed lines with more than two peaks in our simulated spectra (see e.g. panel (c) of Fig.~\ref{fig:B2Tdistribution}), this complexity adds dimensions to parameter space and one would need many more mocks to come close to any such observation. Such unsatisfactory matches thus likely inform us about a limitation of the size of our sample of mock spectra more than on a possible failure of the physical models of our simulation. The presence of a peak at systemic velocity is commonly interpreted as a signature of low-density channels through which \lya{} radiation escapes without scattering much, together with ionising radiation \cite[e.g.][]{Behrens2014a,Duval2014,Verhamme2015,Gronke2016b,Rivera-Thorsen2017}. In that perspective, the mismatch in {\it panels (c)} and {\it (d)} of Fig. \ref{fig:FailedFits} may hint that our simulation does not capture accurately such channels, either because of insufficient resolution or because of the inaccuracies of the moments method used to transport ionising radiation. In our mocks, however, the third peak seems to be an accidental feature, which is rather produced by extra absorption from the extended CGM than by a hole in the ISM that would let \lya{} and Lyman continuum (LyC) photons escape. We have checked that mock \lya{} profiles with multiple peaks indeed do not correspond to situations with larger LyC escape fractions than average, even when selecting triple peaks with a central peak very close to systemic.

\vskip 0.3cm
One of the difficulties of idealised models has been to produce large peak separations with narrow intrinsic lines. \citet{Li2022} have shown how degeneracies in shell models are likely responsible for the excessive intrinsic full width at half maximum (FWHM) of the \lya{} line found by \citet{Orlitova2018} \citep[and before by e.g.][]{Verhamme2008}. They also show how more sophisticated models with a clumpy medium may reproduce observed lines with relatively narrow intrinsic lines. It is indeed important to verify that reproducing the complex \lya{} line is not done at the expense of matching less model-dependent information, e.g. the line width of the H$\beta$ line when accessible. We find that the FWHM\footnote{We define the FWHM as $2.35\times \sigma_v$, where $\sigma_v$ is the flux-weighted velocity dispersion of the intrinsic line. } of the intrinsic \lya{} emission lines in our simulation have a median value of $\sim 120$ km/s, and 80\% of the values are between $\sim 90$ km/s and $\sim 150$ km/s. These values are slightly lower than the H$\beta$ line widths reported by \citet{Orlitova2018}. We note however that these authors derive line widths from relatively low resolution spectra ($\delta v \sim 250$ km/s), which may lead to a systematic overestimation. The intrinsic lines of the best-matching mocks are shown on each panel of Figs. \ref{fig:SuccessfulFits} and \ref{fig:FailedFits} (see also Appendix \ref{app:allfits} and Fig. \ref{fig:B2Tdistribution}). They are always much smaller than the mock-observed line widths, which reminds us that the observed \lya{} line shape is the result of a strong scattering process in the diffusive medium and does not inform us directly on the kinematics of the emitting gas.


\subsection{The distribution of line shapes}\label{sec:LyaShapes}

We have seen in Sec. \ref{sec:LASDComparison} that our mocks provide very satisfactory matches to most observed \lya{} line profiles of low-redshift galaxies, and we have highlighted the minority of cases where no good match is found, mostly in galaxies much more massive than the simulated one, or in extremely transparent (and rare) sight-lines. We now discuss in more detail the distribution of line shapes found in our full sample of mock spectra. 

\vskip 0.3cm

In order to define the line shapes of our 22,500 mock spectra, we first need to detect peaks and valleys in the line profile. Doing this automatically on a large sample of noisy spectra is generally a difficult problem. We are fortunate to have mock spectra with high spectral resolution ($\sim 10$ km/s) and limited Monte Carlo noise, so that a relatively simple method yields robust results. We lay out the details of our empirical method below.

The continuum sometimes has a significant and very broad \lya{} absorption feature around the emission line. In such cases, it is useful to have a model of the continuum, be it approximate, in order to detect weak emission. We model the continuum with a 6th degree polynomial, fit jointly to the red and blue sides of the line: from $-2500$~km/s to $-1500$~km/s on the blue side (to avoid a strong absorption feature which sometimes appears at $\sim -2700$~km/s in the stellar continuum), and from 1500~km/s to 5000~km/s on the red side. While this fit has no physical ground, it provides a smooth continuum model which helps improve the detection of emission peaks and absorption troughs. We use this model to compute a continuum-subtracted signal, which we in turn use to evaluate the noise in our mocks. We do this by measuring the standard deviation $\sigma$ in this signal, over the same velocity interval as was used for the fit, and using 2 iterations of sigma clipping to remove points beyond $3\times \sigma$. We then search for peaks and valleys within $\pm 1500$ km/s of the systemic redshift.  For this, we find that we obtain the best results when smoothing the continuum-subtracted spectra with a Gaussian kernel of width 2 resolution elements ($\sim 20$ km/s). We use the python code {\tt findpeaks}\footnote{This python code was developed by E. Taskesen, based on the 2D implementation of S. Huber \citep{Huber2021}. The code is available at https://github.com/erdogant/findpeaks.} to detect peaks and valleys in these smoothed spectra. This code relies on topological persistence \citep{Edelsbrunner2002} to retain significant peaks and valleys, which we define as those with a persistence score larger than three times the noise level $\sigma$. Finally, we discard peaks with a flux level lower than 3$\sigma$ times the local continuum model, which mostly helps removing false detections far from line centre. We have inspected visually thousands of lines to make sure that our procedure identifies the salient features of the profiles, and yet does not report false peak detections. Indeed, we find that our automatic classification agrees with the eye in all cases we checked. Once peaks and valleys are identified robustly, one can define archetypical line shapes, which we discuss below. 

\vskip 0.3cm 

\begin{figure}
	\includegraphics[width=\columnwidth]{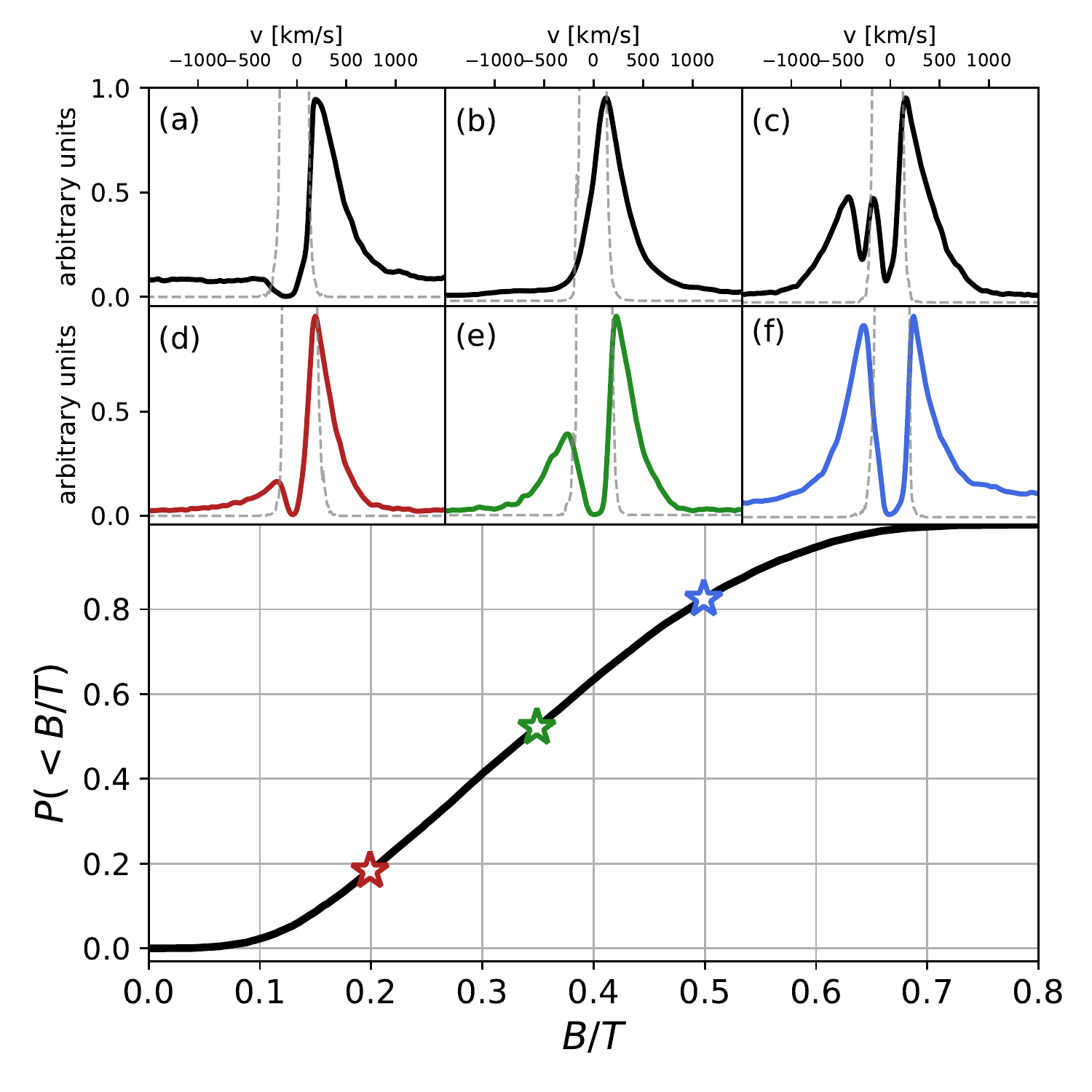}
    \caption{The 6 upper panels, labeled (a) to (f), show selected examples of \lya{} lines from our sample of mock spectra (see text), smoothed with a Gaussian kernel of width 20 km/s. In each of these panels, the thick line is the mock spectrum and the thin dashed grey line is the intrinsic emission profile in the same direction, using the same normalisation. The lower panel shows the cumulative distribution of blue-peak to total flux ratio ($B/T$) in our sample of double-peak profiles. The three coloured stars indicate $B/T$ values corresponding to panels (d), (e), and (f). }
    \label{fig:B2Tdistribution}
\end{figure}

{\bf P-Cygni profiles} are those that have one single peak and one valley detected. Panel (a) of Fig. \ref{fig:B2Tdistribution} shows such a P-Cygni profile from our sample, with saturated absorption on the blue side and a strong, asymetric peak on the red side. Such profiles are commonly observed in low-redshift galaxies \citep[e.g.][]{Wofford2013,Rivera-Thorsen2015} and in the distant Universe \citep[e.g.][]{Shapley2003,Inami2017,Oyarzun2017}. While at high redshifts, scattering through the intergalactic medium plays a role in shaping such lines \citep[see][]{Laursen2011,Garel2021}, their observation at low redshifts plainly shows that the ISM and CGM alone may produce them. So far, simulations of high-redshift galaxies have managed to reproduce approximately such line shapes only thanks to strong intergalactic absorption \citep[see for example the recent works by][]{Smith2019,Behrens2019}. The mocks presented here are the first occurrence, to our knowledge, of P-Cygni profiles from a simulated galaxy without any attenuation from the IGM. These profiles are nevertheless a minority in our sample, where they represent about 1\% of the lines. Note that the peaks of all the P-Cygni profiles we find in our mocks are redshifted relative to systemic. We find the absorption trough most often peaks bluewards of systemic, but is occasionally also a bit redshifted.

\vskip 0.2cm
{\bf Single peaks} are lines where we detect a single peak and no valley. Panel (b) shows an example. Such lines are observed at low redshift \citep[e.g.][]{Yang2017}, and at high redshifts \citep[e.g.][]{Cao2020}. Note that single peak profiles come with many shapes: some are red with a sharp drop towards the blue, similar to P-Cygni profiles, while some show a strong and extended excess of flux on the blue side of the peak, very similar to double peaks, but without any absorption trough. These profiles make about 4\% of the line profiles of our full sample. 

\vskip 0.2cm
{\bf Multiple peaks} are line profiles where more than two peaks (and more than one valley) are detected. Panel (c) shows an example of such a line profile from our sample. Such multiple peak profiles are observed, albeit rarely, and it is interesting that we also find these odd cases in our sample. Galaxies J0007+0226 or J1032+4919 from \citet[][]{Izotov2020} are examples of such complex line profiles, which are generally not reproduced with idealised models \citep[][appendix B]{Gurung-Lopez2022}, even if they are sometimes found to emerge from multiphase media \citep{Gronke2016}. Regardless of the weak peak in the centre, this example also illustrates the possibility to find significant residual flux between the red and blue peaks in our mock spectra. These peculiar line profiles represent about 5\% of our mocks. 

\vskip 0.2cm
{\bf Double peaks} are profiles where we detect two peaks (and most often one single valley). Panels (d) to (f) show double peak profiles with an increasingly strong blue peak. Panels (d) and (e) are the more archetypical \lya{} profiles, with a dominant asymmetric red peak and an absorption trough roughly at systemic velocity. Such lines are often seen at low redshifts \citep[e.g.][]{Henry2015} or high redshifts \citep[e.g.][]{Tapken2007,Trainor2015,Cao2020}. Panel (f) shows a line with equal flux in the red and blue peaks, both falling sharply towards zero-velocity, where the absorption saturates. Such lines are observed as well, as reported for example by \citet{Tapken2004} or \citet{Kulas2012} at high redshifts. Our sample also contains lines where the blue peak dominates and is broader than the red peak. While dominant blue peaks seem to be rare in nature, some are nevertheless observed in the local Universe \citep[e.g.][]{Wofford2013} and even at higher redshifts despite the effect of the IGM \citep[e.g.][]{Kulas2012,Furtak2022,Marques-Chaves2022}. Some of these broad double peak profiles are very hard to reproduce with shell models without unrealistically broad intrinsic emission \citep[e.g.][]{Orlitova2018,Gurung-Lopez2022}. The fact that the intrinsic emission lines are here always much thinner than the observed ones tells us that scattering operates in our simulations in a way which may be closer to reality than can be achieved with idealistic models.

\vskip 0.2cm
Double peak profiles are the most common in our sample and make up 90\% of the lines. It is perhaps not surprising that we find a majority of double peaks, because we can detect very faint blue bumps barely above continuum level. The observed fraction of double peaks will however be affected by spectral resolution and noise, and we expect it to be lower. Interestingly, the fraction of lines with 3 peaks or more is low and not limited by noise or resolution, and we find no more than a handful of profiles with more than 4 peaks or more among our 22,500 mocks. Similarly, single peaks and profiles with very extended signal to the blue but no absorption trough are genuine. 

\begin{figure*}
	\includegraphics[width=\textwidth]{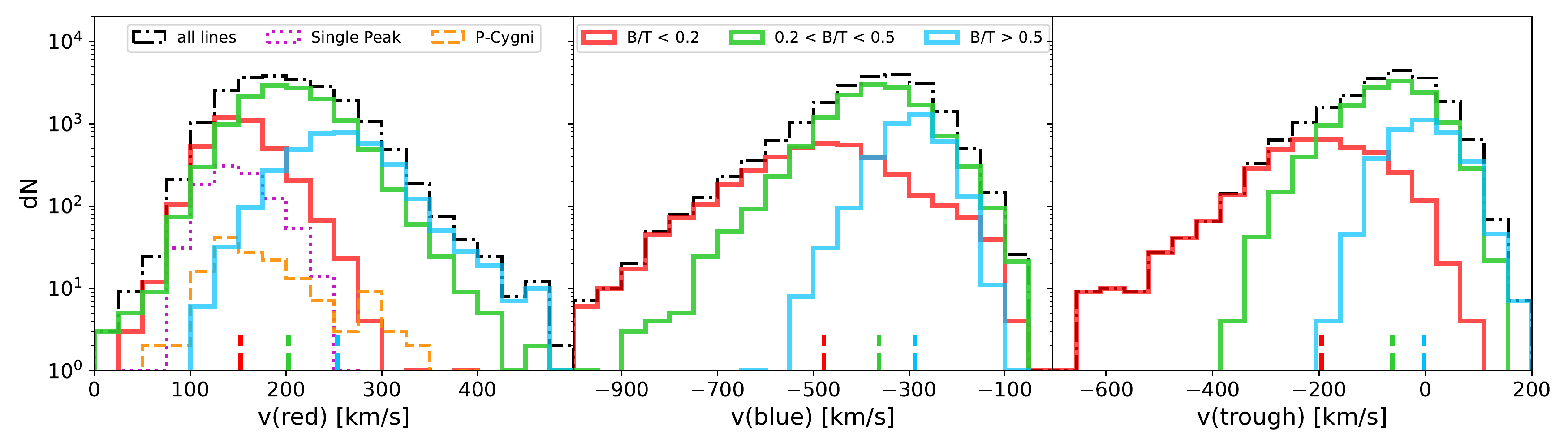}
    \caption{Distributions of peak and trough velocities, relative to systemic, of mock profiles. Panels from left to right show the red peak, blue peak, and trough velocity distributions. The black dash-dotted line shows the distribution of all mock \lya{} lines on each panel. The red, green, and blue lines show the distributions for double peak profiles with blue-to-total flux ratios $B/T < 0.2$, $0.2< B/T < 0.5$, and $B/T>0.5$ respectively. In the left panel, we also show the distribution of P-Cygni and single peak profiles with dotted and dashed lines. The median values of all distributions are indicated by short vertical lines at the bottom of each panel, except for P-Cygni and single peak profiles which have a very similar median peak velocity to the reddest double peaks. }  
\label{fig:vpeak}
\end{figure*}

\vskip 0.2cm
The large population of double peaks hosts a diversity of line shapes, with very different relative fluxes in the blue and red peaks. We measure these relative fluxes using the blue-to-total flux ratio $B/T = F_{\rm blue} / (F_{\rm red}+F_{\rm blue})$, where $F_{\rm blue}$ and $F_{\rm red}$ are the fluxes in the blue and red peaks\footnote{We compute peak fluxes by integrating the spectrum left and right of each peak location, down to the velocities where the spectrum reaches 1 $\sigma$ above the continuum model, or down to the valley between peaks if it is above the continuum level.}. The lower panel of Fig. \ref{fig:B2Tdistribution} shows the cumulative distribution of $B/T$, using all double peak profiles in our sample of mocks. There are a number of important points to take from this distribution. 
First, {\it mock \lya{} lines are generally red}: more than 80\% of our mock double peaks have more than half the flux in the red peak. Less than 20\% of the lines have more flux in the blue peak than in the red peak, and almost none have a blue peak with more than twice the flux of the red peak (i.e. $B/T > 2/3$). In symmetry, 40\% of our double peak profiles have more than 70\% of the flux in the red peak. While this distribution is not directly comparable to observations (see Sec. \ref{sec:discussion}), it is clear that the observed \lya{} lines follow a similar pattern and generally display a stronger red peak. The fact that most of our \lya{} lines are red is an important success of our simulation in that sense, which contrasts with previous numerical works \citep{Tasitsiomi2006a,Laursen2007,Laursen2009a,Laursen2009,Verhamme2012,Yajima2012,Behrens2014,Behrens2019,Smith2019,Garel2021,Smith2022b,Smith2022a}. 
Second, we find a relatively significant population of double peaks with very blue ($B/T > 0.5$) or very red ($B/T < 0.2$) profiles, with a bit less than 20\% of the population in each extreme. The bulk (60\%) of our mock double-peak profiles are more ordinary double peaks with $0.2 < B/T< 0.5$, i.e. between 50 and 80\% of the line flux in the red peak.


\subsection{Line parameters}\label{sec:LineParams}

The blue-to-total flux ratio is only one of the parameters that defines a double peak profile, and the examples shown in the upper panels of Fig. \ref{fig:B2Tdistribution} do not give a full impression of the diversity of double peaks that we find in our sample. Indeed, we have seen in Sec. \ref{sec:LASDComparison} that some lines from the LASD are not well matched by our mocks, and it is useful to understand more generally how our mock lines cover parameter space.

In Fig. \ref{fig:vpeak}, we show the distributions of peak and trough velocities of our mock \lya{} lines. Looking first at the dot-dashed black lines in each panel, we see that our sample of mock spectra covers a large range of peaks and trough positions. The red peaks in our mocks are found at velocities distributed around a median value of $\sim 200$~km/s, with only 10\% of lines peaking below 140 km/s and 10\% above 270 km/s. Note that we count P-Cygni and single peaks as red peaks here, which does not change the results significantly. Similarly, 80\% of our blue peaks are found at velocities between $-520$ and $-250$~km/s, with a median velocity of $\approx -360$~km/s. Our double peak profiles have absorption troughs which tend to be on the blue side, with a median velocity of $\approx -70$~km/s and 80\% of the mocks between $-215$ km/s and $30$~km/s. These values broadly cover the range of values measured in the spectra of the LASD, and it is interesting that the blue or red peak velocities generally largely exceed the circular velocity of the dark matter halo, which is only about 90~km/s. This again underlines the important role of scattering in shaping the observable properties of the \lya{} line. 

There are a few additional features to note in Fig. \ref{fig:vpeak}. First, there is an asymmetry in the velocity distribution of blue and red peaks: blue peaks are bluer than red peaks are red. As we will see in Sec. \ref{sec:cgm}, this is mostly due to the outflowing CGM producing an effective absorption trough on the blue side of the line, which pushes the position of the blue peak to bluer velocities. Indeed, both the trough velocity and the blue peak velocity move to more negative velocities when the lines become redder (i.e. when $B/T$ decreases). Second, blue and red peaks close to systemic are rare, and it is thus not surprising that we have difficulties reproducing narrow lines from the LASD: our sample does not contain enough of these. At this point, it is unclear whether this is a limitation of the physics in the simulation or strong selection effects on the observational side and a lack of statistics on the simulation side: strong LyC leakers are rare and their \lya{} lines are peculiar. Third, the absorption valley is generally in the blue, although a non-negligible fraction of mocks have a trough peaking at positive velocities. The common interpretation of a blue-shifted absorption is that radiation scatters through an outflowing medium, and we will come back to this in Sec. \ref{sec:SightLineVariations}. 

\begin{figure}
	\includegraphics[width=\columnwidth]{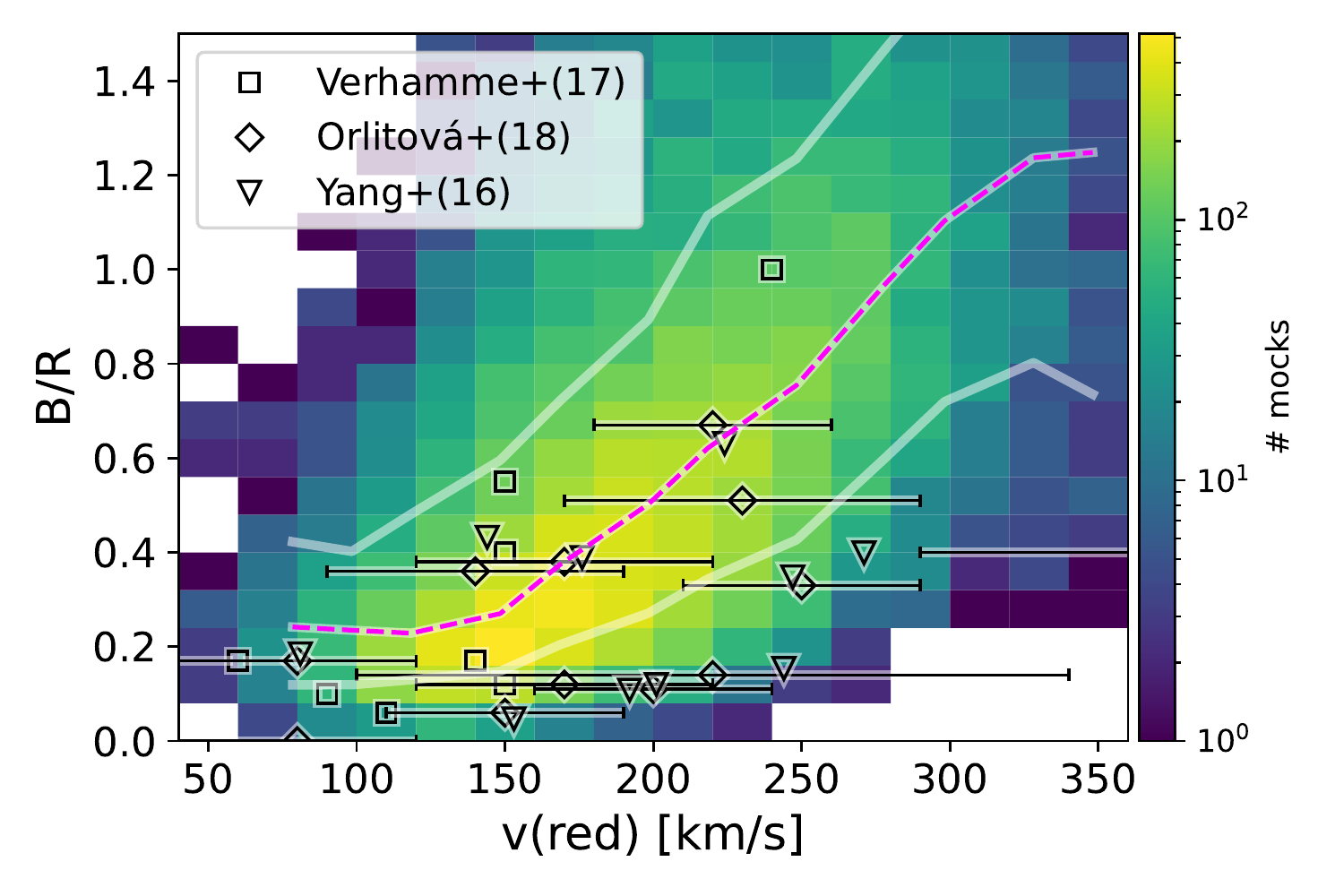}
    \caption{Distribution of the blue-to-red flux ratio as a function of the red peak velocity for all double peaks in our mocks. The dashed magenta line shows a running median in bins of red-peak velocity, and the white lines above and below indicate the 15\% and 85\% percentiles. The data points with error-bars are observational estimates from the literature as indicated on the plot. }
    \label{fig:b2t_vs_vred}
\end{figure}

Fig. \ref{fig:vpeak} also presents the distributions of peak velocities for different classes of line profiles, and shows very strong variations in these distributions depending on the blue-to-total flux ratio. Very blue double peak profiles ($B/T > 0.5$), which are generally interpreted as a sign of inflow \citep{Dijkstra2006}, are the most symmetric, with an absorption close to systemic and median peak velocities $v_{\rm red}\sim 250$ km/s and $v_{\rm blue}\sim -300$ km/s. The reddest profiles, on the contrary, have the lowest red peak velocity ($v_{\rm red}\sim 150$ km/s), and blue-shifted absorptions which push the blue peaks to very negative velocities. The bulk of the population, with intermediate $B/T$ ratios, populates a parameter space between these two extremes. \citet{Kimm2022} carried out sub-parsec resolution simulations of turbulent molecular clouds, from which they extracted mock \lya{} spectra. In their Fig.~10, they compare their results to a compilation of observations in the plane of blue-to-red flux ratio versus red-peak velocity. Their \lya{} lines span a broad range of blue-to-red flux ratios, but tend to pile up at relatively low velocities ($v_{\rm red}\sim 50$ km/s), except for short excursions in the early phases of a cloud's evolution. The origin of this difficulty to produce large peak velocities may well be due to the fact that their mocks do not account for scattering through the more diffuse ISM and through the CGM, which are not present in their molecular cloud simulations. In Fig.~\ref{fig:b2t_vs_vred}, we show that the situation is quite different for our mock observations. This figure shows in a more quantitative manner than Fig.~\ref{fig:vpeak} the correlation between the colour of the line (measured by the $B/R$ flux ratio) and the red peak velocity: redder lines (i.e. lower $B/R$ values) tend to have lower velocity offsets than bluer lines. Interestingly, this trend seems to be compatible with observations of local \lya{} lines. As we will see in Sec. \ref{sec:cgmspec}, we can interpret this as being due to a combination of lines generally emerging from the galaxy with a red peak at $\sim 150$~km/s, and of the effective absorption by the CGM which is blue-shifted for very red lines, but closer to systemic for very blue lines. This broad CGM absorption close to systemic at large $B/T$ values extends into the red peak and pushes it to larger velocities. 

\begin{figure}
	\includegraphics[width=\columnwidth]{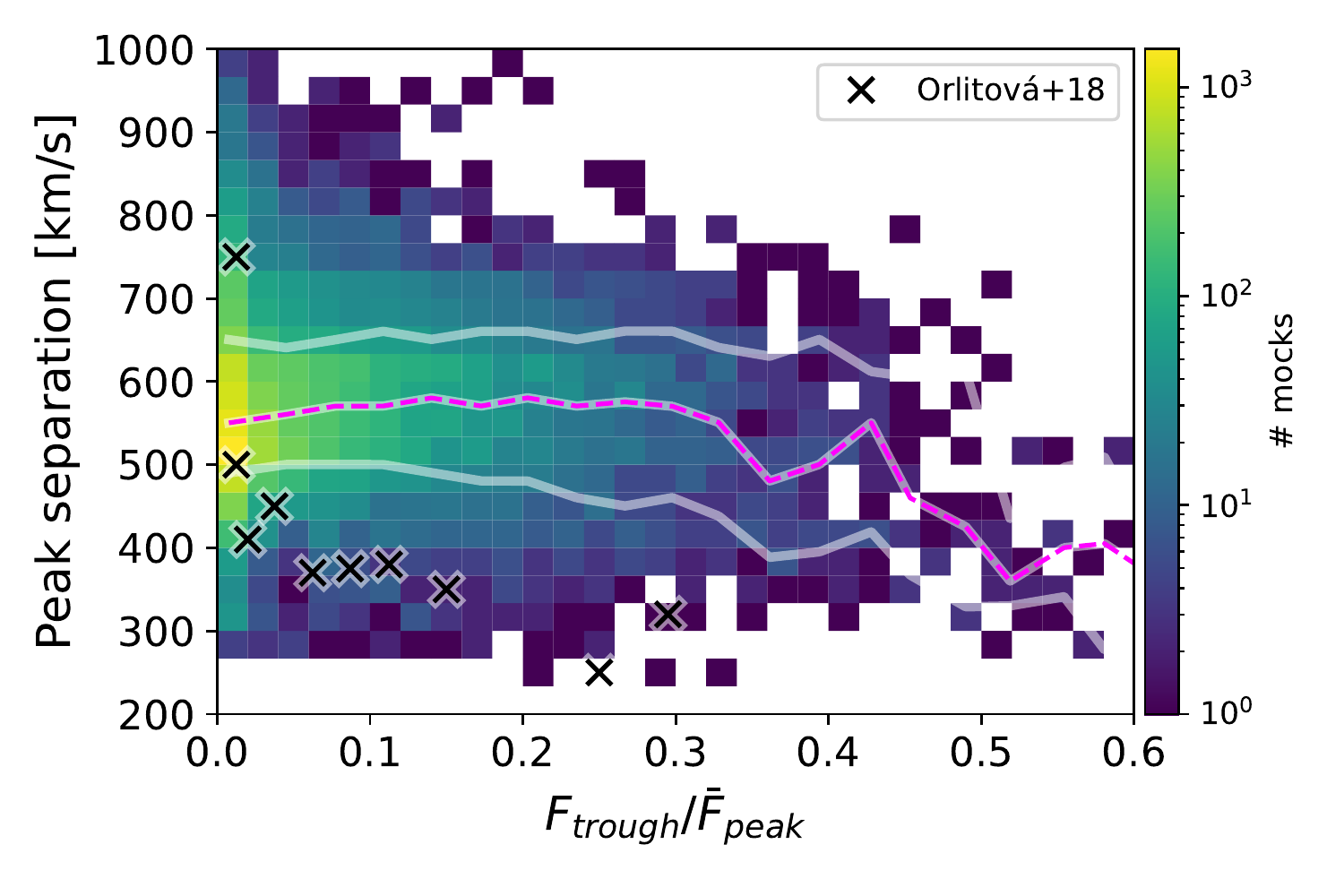}
    \caption{Distribution of our double peak profiles with $B/T>0.2$ in terms of peak separation and trough to mean-peak flux ratio. The black crosses are points taken from \citet{Orlitova2018}. The dashed magenta line shows the running median of the mock double peaks, and the white lines above and below indicate the 15\% and 85\% percentiles.}
    \label{fig:Orlinke}
\end{figure}

Another key parameter of double peak profiles is the peak separation $v_{\rm sep}=v_{\rm red} - v_{\rm blue}$. Scattering through higher column densities of H{\sc i} will increase peak separation, and indeed $v_{\rm sep}$ has been shown to correlate at least loosely with the escape fraction of ionising radiation \citep{Verhamme2015,Verhamme2017,Izotov2018,Steidel2018}. Another quantity which is directly related to the H{\sc i} column density in idealised models is the residual flux level in the absorption trough, $F_{\rm trough}$. \citet{Gronke2016} suggested that the ratio of $F_{\rm trough}$ to the mean flux in the peaks ($\bar{F}_{\rm peak}$) may help discriminate models, as it has a different distribution for shell or clumpy models. In Fig. \ref{fig:Orlinke}, we show the distribution of our mock spectra in the $v_{\rm sep}-F_{\rm trough}/\bar{F}_{\rm peak}$ plane. Also shown are values reported by \citet{Orlitova2018} for a small sample of green pea galaxies. In both properties, our mocks cover the observed range reported by \citet{Orlitova2018} and beyond. Interestingly, our mocks produce a distribution of $F_{\rm trough}/\bar{F}_{\rm peak}$ which lies somewhere in between the extreme examples examined by \citet{Gronke2016}. Most of our galaxies have a strong absorption, with $F_{\rm trough}/\bar{F}_{\rm peak}\sim0$, as for shell models. Yet, we do find a significant population with positive residual flux, which these authors find only for their clumpy models. In terms of peak separation, our mocks do not produce as large separations as the grid of models of \citet{Gronke2016}, which may again be due to the fact that our simulated galaxy is a low-mass LAE. The mocks with large residual fluxes are intriguing and we have searched whether they correspond to sight-lines with more or less H{\sc i} content or larger turbulent velocities. We find no obvious correlation between the residual flux and these quantities, and these lines are likely the product of more subtle effects.

\begin{figure}
	\includegraphics[width=\columnwidth]{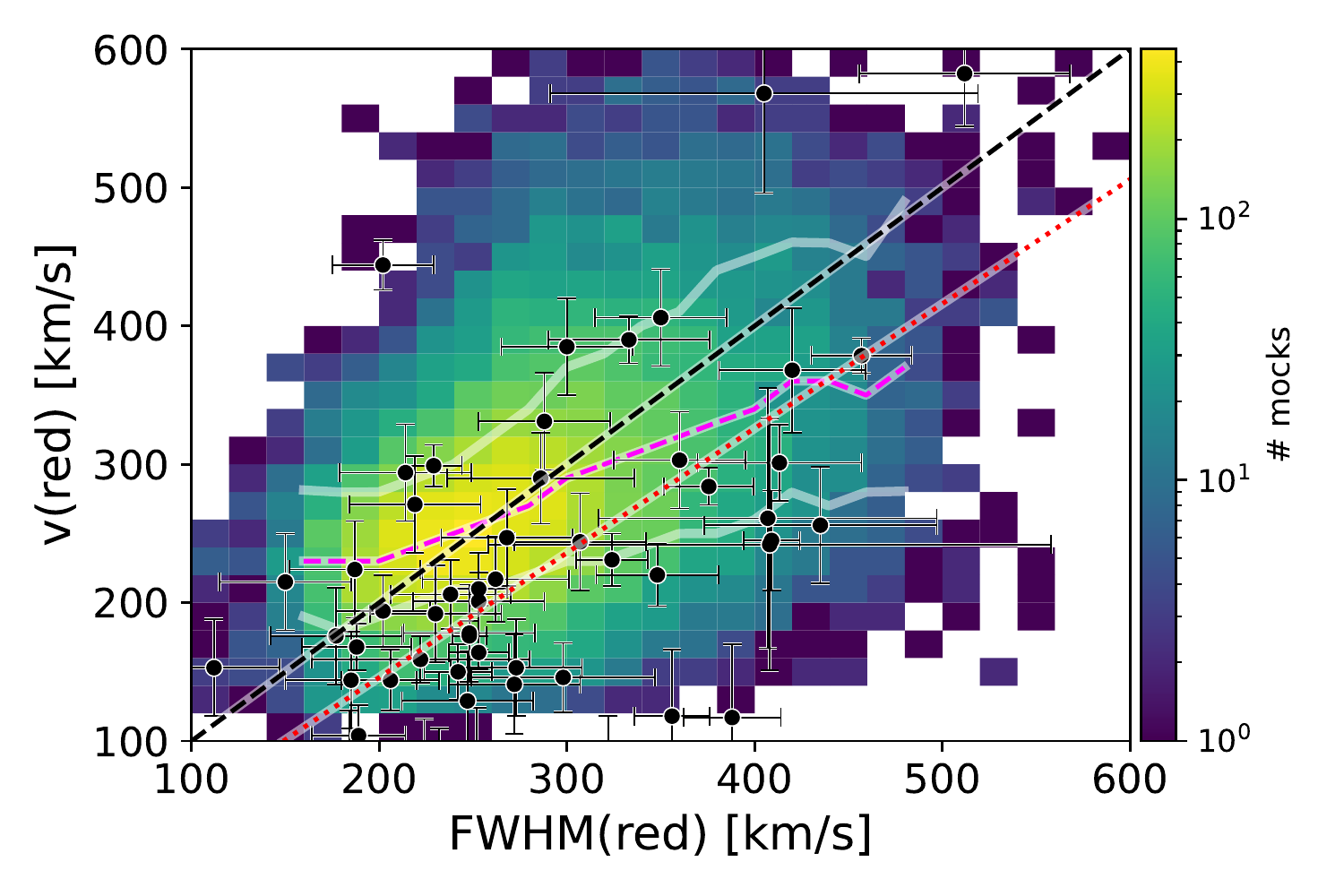}
    \caption{Distribution of red peak velocity and width for all our mocks. The dashed black curve shows a one-to-one relation, and the red dotted line is the best-fit relation from \citet{Verhamme2018}. The symbols with error bars are the observational points compiled by \citet{Verhamme2018}. The dashed magenta line shows the running median of the mock double peaks, and the white lines above and below indicate the 15\% and 85\% percentiles.}
    \label{fig:vred_vs_fwhm}
\end{figure}

As we saw in Sec. \ref{sec:LASDComparison}, our mocks do not reproduce the very narrow \lya{} lines observed for Lyman continuum leakers. Indeed, the width of the \lya{} line is yet another key shape parameter which is often measured and may contain information on the physical processes that shaped the \lya{} line. In Fig. \ref{fig:vred_vs_fwhm}, we show the distribution of the peak velocity versus line width for the red peaks of our mock spectra (including P-Cygni and single peak profiles). All line types have similar distributions, with a median FWHM$\sim 280$ km/s. Interestingly, we find a relatively shallow relation between the peak velocity and its FWHM, and with significant scatter. In Fig. \ref{fig:vred_vs_fwhm}, we also report the best-fit relation from \citet{Verhamme2018}, and the one-to-one relation, which are both steeper than our distribution. It is important to have in mind that the observed relations have been derived for samples of galaxies with complex selection functions, and that our mocks are selected in a very different way. With Fig. \ref{fig:vred_vs_fwhm} (and Figs. \ref{fig:b2t_vs_vred} and \ref{fig:Orlinke}), we aim to show that the mocks cover a similar part of parameter space as observations, but we should not push the statistical comparison too far (see Sec. \ref{sec:caution}). 

\vskip 0.3cm 
We conclude this section by underlining that the sample of mock spectra we have constructed from {\it the simulation of a single high-redshift galaxy} covers a significant part of the diversity of line shapes which are observed at low and high redshifts. As we have shown in Sec.~\ref{sec:LASDComparison} and App. \ref{app:allfits}, our mock lines match accurately most of the spectra available in the Lyman-Alpha Spectral Database. One galaxy cannot fit them all, though, and we find two main populations missing from our sample: (1) The \lya{} lines from more massive galaxies, which feature very broad and deep absorption features, are not well matched by our sample; (2) The peculiar \lya{} lines of some extreme Lyman continuum leakers are not found, possibly because they are too rare. We have seen in Sec.~\ref{sec:LyaShapes} that our low-mass galaxy produces mostly double peak profiles, which are generally red, as in observations. We have seen that this line shape is the result of strong scattering effects, as the observed line is always much broader than the intrinsic emission line. Finally, in Sec.~\ref{sec:LineParams}, we have shown that the parameters which characterise the \lya{} line shape cover a large range of values, comparable to observations. We have also seen that despite the fact that we mock-observe a single galaxy, some observational trends seem to emerge in the sample of mocks, e.g. relating the red peak velocity to the blue-to-red flux ratio or to the FWHM of the red peak.

\section{What determines the \lya{} line shape} \label{sec:SightLineVariations}

In Sec.~\ref{sec:ManyShapes} we have shown that our simulated spectra qualitatively reproduce the diversity of observed line shapes, and are indeed able to reproduce very satisfactorily the majority of observed low-redshift spectra. It is remarkable that this success is obtained by mock-observing {\it a single simulated galaxy evolving over $\sim700$Myr}, and we now investigate the origin of the many line shapes produced by this single object. 

\subsection{Variation of the line shape with direction}\label{sec:varDir}

\begin{figure}
\includegraphics[width=\columnwidth]{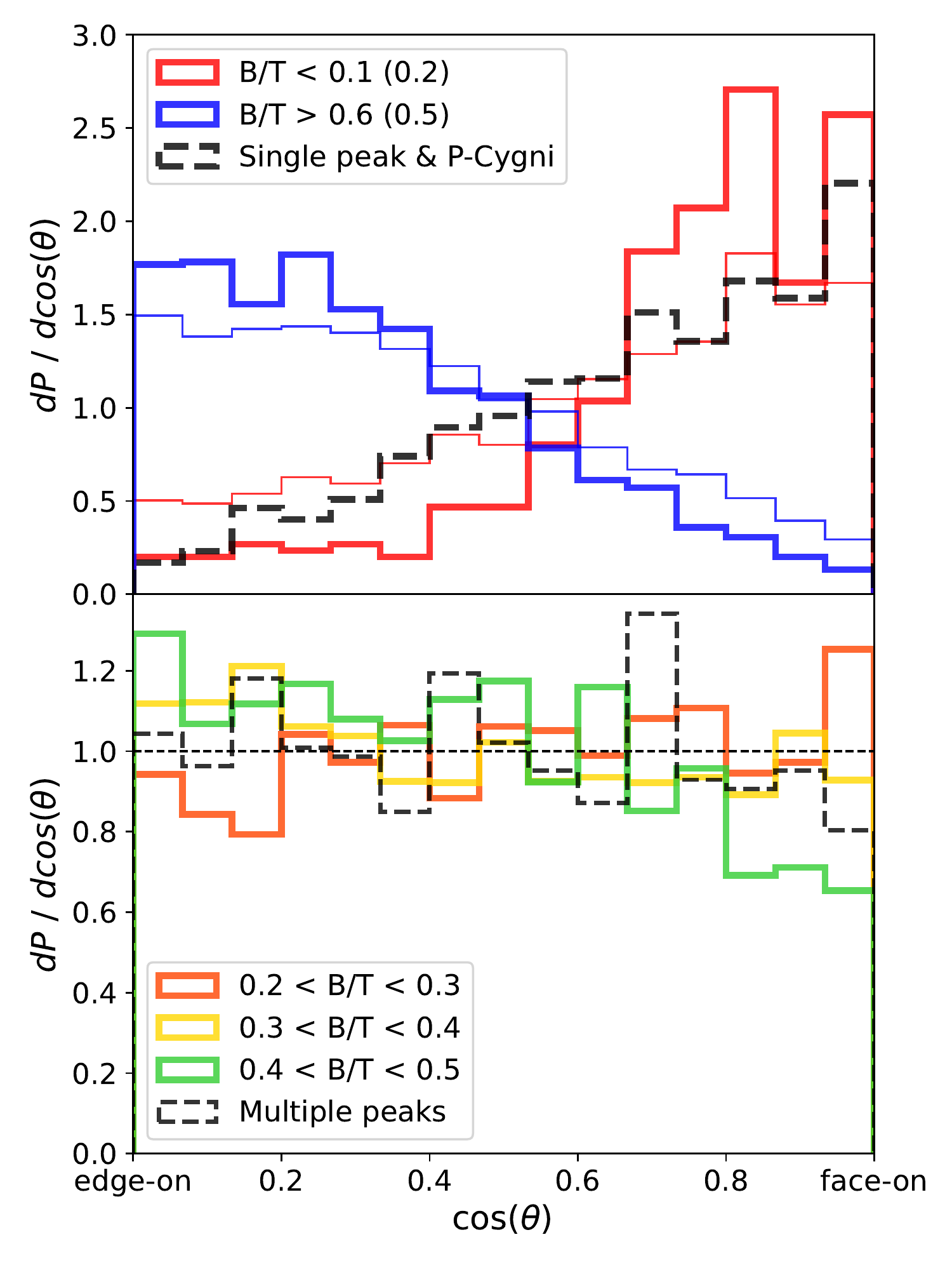}
\caption{Probability density function of the inclination at which different line types are observed. These PDFs are computed using all 22,500 mock spectra and shown in two panels for better readability. The upper panel shows the PDFs of extremely blue and red double peak profiles, along with P-Cygni and single peaks. The lower panel shows the PDFs of ordinary double peak profiles with $B/T$ between 0.2 and 0.5, as well as the PDF of multiple peak profiles which is similar. The normalisation of the PDFs is such that a constant value of unity means isotropy. In the upper panel, the thin lines correspond to $B/T$ values indicated in parenthesis in the legend. }
\label{fig:LineTypes_vs_direction}
\end{figure}

The variations of \lya{} properties with sight-line have been reported in the context of idealised disc-galaxy simulations \citep{Verhamme2012,Behrens2014,Smith2022b}. While these works reach a general consensus for example on the variation of \lya{} escape fraction with inclination, they disagree on the variations of the line profile with inclination. \citet[][their Fig. 5]{Verhamme2012} find that symmetric double peak profiles are produced edge-on, while face-on line profiles are strongly asymmetric with a dominant red peak. \citet[][their Fig. 2]{Behrens2014} find a relatively symmetric double peak regardless inclination, and \citet[][their Fig. 16]{Smith2022b} find that double peaks are seen face-on, while edge-on mock observations produce extended single peak profiles.  

Numerical studies using cosmological simulations have also investigated how the \lya{} line shape may vary with direction, albeit with less detail \citep{Laursen2007, Behrens2019, Smith2019, Smith2022a}. In particular, \citet{Smith2019} show that the line properties are very anisotropic and show hints of correlations with spatially resolved motions of the gas in and around their simulated galaxy. \citet{Smith2022a} extend this analysis and show that the small-scale effects that are responsible for the line shape are correlated, at high redshifts, to the transmission of the IGM. These analysis are however limited by the fact that none produce line shapes that compare well to observations without the help of ad-hoc sub-grid modelling in post-processing, or the effect of the IGM. 

Given the new degree of realism achieved by our simulated spectra, it is worth revisiting this question in more detail. In Fig.~\ref{fig:LineTypes_vs_direction}, we show the probability density function (PDF) of inclinations at which different line shapes are observed\footnote{We define the inclination relative to the angular momentum vector of the gas within $0.2 \times R_{vir}$ of the halo centre. We have carried out the same analysis using the angular momentum of stars and find very similar results, albeit somewhat noisier.}. This PDF shows the relative likelihood that a line shape appears at some inclination rather than another, and is normalised so that a constant value of one means isotropy. In the upper panel, the thick blue (red) curve shows the PDF of double peak profiles with more than 60\% (less than 10\%) of the flux in the blue peak. These extreme line profiles show a clear trend with inclination: very blue lines are observed preferentially edge-on, whereas very red lines are rather seen face-on. This trend holds when we relax slightly the selection criterion and consider blue profiles with $B/T > 0.5$ (thin blue curve) or red profiles with $B/T < 0.2$ (thin red line). As expected, the PDFs of single peaks and P-Cygni profiles follow that of red lines. Quantitatively, when a blue-dominated line is observed, it is 5-12 times more likely to be seen edge-on than face-on. Conversely, when a very red profile is observed, it is 5-14 times more likely to be face-on than edge-on. These results align well with the behaviour noted by \citet{Verhamme2012}, although they differ in the details, but are rather orthogonal to the results of \citet{Smith2022b}. 

The lower panel of Fig.~\ref{fig:LineTypes_vs_direction} shows that the picture is more subtle. There, we see that the PDFs corresponding to the most common double peak profiles, with a blue fraction ranging from 20 to 50\%, are roughly flat and equal to unity, except for a shallow trend for the bluer lines. This is also the case for the PDF of profiles with multiple peaks. In other words, these line profiles are seen in any direction with no preference, and observing such a line does not inform us on the inclination of the galaxy. From Sec.~\ref{sec:LyaShapes}, we know that these profiles are common in our sample, where they represent about 60\% of all mocks. Thus, in our sample, observing an extremely red line means we are likely seeing the galaxy face-on, but face-on mocks nevertheless mostly produce rather ordinary double peak profiles. Similarly, extremely blue lines are preferentially seen edge-on, but edge-on mocks mostly produce ordinary double peak profiles. With JWST, it has certainly become possible to test for such correlations between the \lya{} line shape and the inclination of a galaxy. Such tests would bring stringent constraint on theory and help discriminate models. 

These results produce a more nuanced relation between line shape and inclination than the one obtained with idealised disc setups. This is expected because of the irregular morphology of our galaxy and its structured CGM. Still, the scattering process outlined by \citet{Verhamme2012} seems to hold, as we will confirm in Sec. \ref{sec:cgm}.

\subsection{Variation of the line shape with time} \label{sec:varTime}

\begin{figure}
\includegraphics[width=\columnwidth]{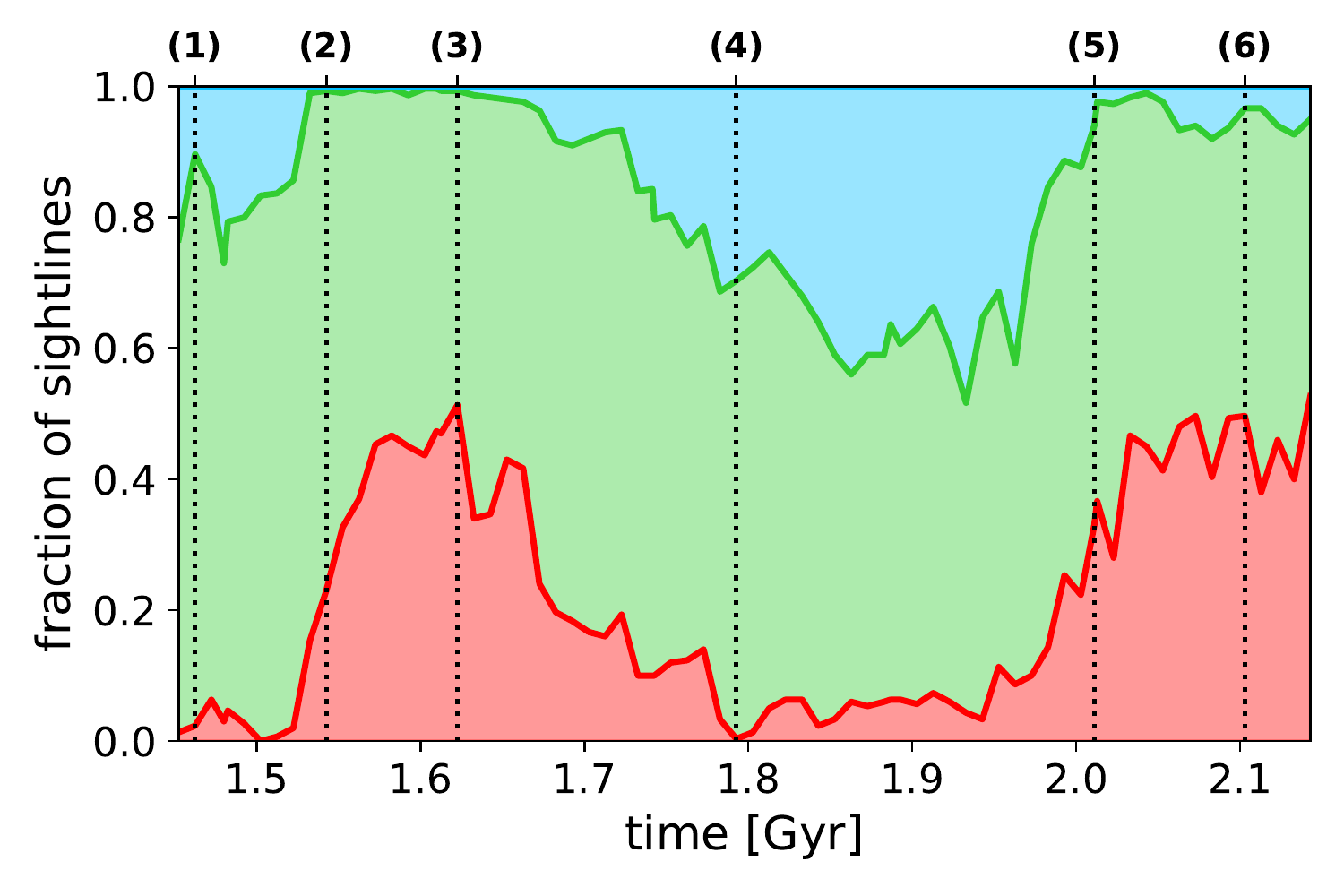}
\caption{Distribution \lya{} line types as a function of time. The red curve (and red shaded area) shows the fraction of sight-lines with very red line profiles: P-Cygny, single peaks, or double peaks with with $B/T < 0.2$. The green shaded area shows the fraction of ordinary double peaks ($0.2< B/T < 0.5$) and multiple peaks. The blue shaded area shows the fraction of very blue profiles ($B/T > 0.5$). The vertical dotted lines and labels above the plot indicate particular times which are illustrated in Fig.~\ref{fig:sequence}.}
\label{fig:LineTypes_vs_time}
\end{figure}

We now turn to variations of the distribution of line shapes with time, and first define three line types for simplicity. {\it Very red lines} have either a single peak (this includes P-Cygni profiles) or two peaks with $B/T < 0.2$. {\it Ordinary double peaks} have two peaks with $0.2 < B/T < 0.5$. We also include multiple peaks in this category, because these can most often be interpreted as a double peak with an extra absorption feature, and we have shown in the previous section that they have a similar behaviour as ordinary double peaks in terms of angular distribution. Finally, {\it very blue lines} are double peaks with $B/T \geq 0.5$. We remind the reader that in our sample, this type of lines mostly include double peaks with similar or slightly more flux (up to a factor 2) in the blue peak than in the red (see Fig.~\ref{fig:B2Tdistribution}). These lines are thus {\it very blue} relatively to the sample, but they remain typical \lya{} lines compatible with observations. 

In Fig.~\ref{fig:LineTypes_vs_time}, we show the fraction of mock sight-lines with such line types as a function of time. A first point to take from this figure is that the joint population of very red and ordinary lines account for a majority of profiles at all times. We already saw in Sec.~\ref{sec:LyaShapes} that most of our mock spectra produce red \lya{} lines, and Fig.~\ref{fig:LineTypes_vs_time} further shows that this is true at any time. The fraction of very blue lines varies strongly in time, but is always smaller than $\sim 40$\% and vanishes at times. The fraction of very red lines also varies strongly, from zero to half the sight-lines. Very interestingly, the fractions of very blue and very red lines evolve in phase opposition: when one is highest the other is lowest, and it thus seems that our simulated galaxy generally does not produce both these extreme line profiles at the same time. The timescale over which these fractions vary is $\sim 200$ Myr, which allows for a few oscillations over the period during which we construct mock observations. It is clear that there is a correlation between these oscillations and the star formation history of the galaxy (Fig. \ref{fig:SFR_Mstar}), and we will see below that these fluctuations correspond to a cycle of accretion and ejection of gas from the simulated galaxy. The extreme line shapes thus seem to inform us, when observed, on the dynamics of the CGM. 

Another important feature of Fig.~\ref{fig:LineTypes_vs_time} is that the fraction of ordinary double peak profiles is roughly constant. At any time, the galaxy will be seen as an ordinary double peak in $\sim40-60$\% of the directions. If anything, this suggests that ordinary double peak profiles do not tell us much about the instantaneous state of the gas in and around the observed galaxy, as they do not inform us about inclination. They are always there with similar properties and probability regardless of the evolutionary phase of the galaxy. 

\begin{figure*}
	\includegraphics[width=\textwidth]{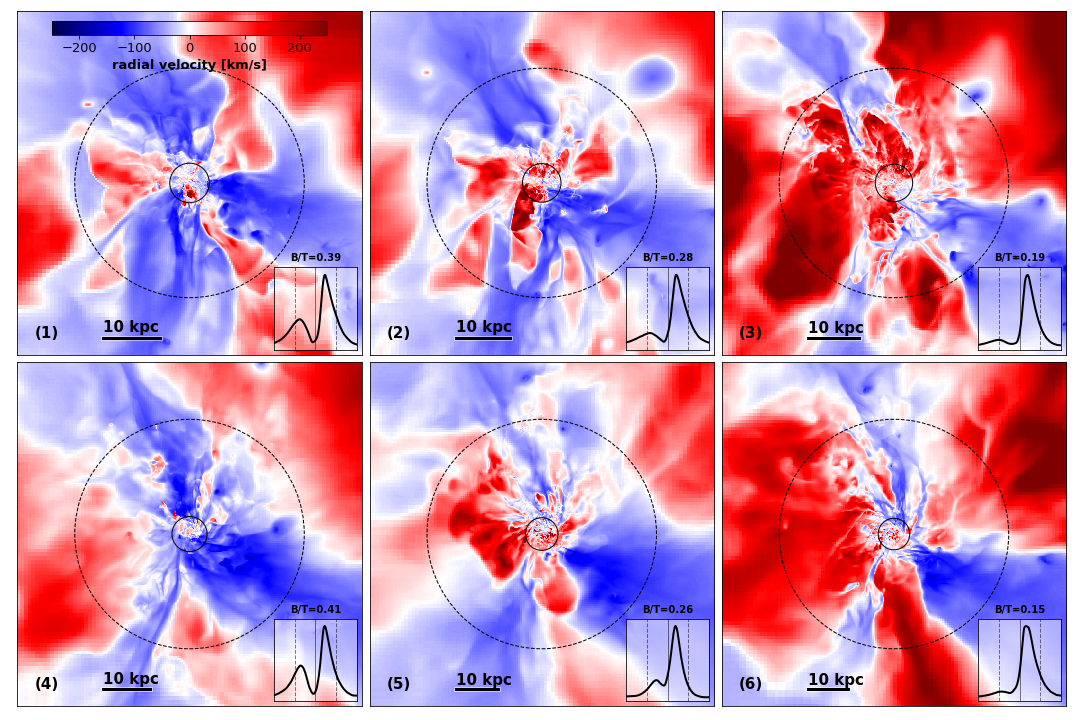}
    \caption{Maps of density-weighted mean radial velocity along the line of sight in and around the simulated galaxy at the 6 snapshots highlighted in Fig.~\ref{fig:LineTypes_vs_time}. In each panel, the inner solid-line circle shows the 1 arcsec aperture within which the spectra are collected, and the dashed-line circle indicates the virial sphere. The horizontal bar in the bottom of each panel shows a scale of 10 proper kpc. All panels share the same colorbar, which is shown in panel (1). The insets in each panel show the normalised angle-averaged spectra corresponding to each time. The $x$-axis of these insets runs from $-1000$~km/s to $1000$~km/s, the solid vertical line marks the systemic velocity and the dashed lines indicate $\pm 500$~km/s. The text above each inset gives the $B/T$ value of the angle average spectrum. }  
\label{fig:sequence}
\end{figure*}

\subsection{A sequence of inflows and outflows}\label{sec:timesequence}

The marked dependence of extreme line types on orientation and time suggests an association with inflows and outflows of gas in the CGM. Indeed, we do find such a connection, which we illustrate in Fig.~\ref{fig:sequence} with radial velocity maps of gas in and around the simulated galaxy at the different times highlighted in Fig.~\ref{fig:LineTypes_vs_time}. In the upper left panel of Fig.~\ref{fig:sequence}, we see an initial phase, close to $z=4$, where three broad, volume-filling, streams of relatively cold gas fall onto the galaxy. At this time, the galaxy produces mostly ordinary double peak profiles and a small fraction ($\sim 10\%$) of very blue lines. These streams bring gas into the central galaxy and feed star formation, which soon triggers a violent outflow. Panel (2) shows the onset of this outflow, when the fast wind has already broken out of the galaxy and reached 20-50\% of the virial radius. Note that the gas within the 1 arcsec aperture ($\sim 0.15\times R_{\rm vir}$) is also expanding: the galaxy is blowing out completely. At this time, the galaxy has begun to produce very red line profiles in $\sim 20\%$ directions. This fraction rises to $\sim$50\% as the outflow further develops at all scales, and we see in panel (3) a situation where most of the gas in the galaxy and in the halo (and beyond) is outflowing at large velocities ($\gtrsim 200$ km/s). The cold streams are significantly affected by the strong galactic outflow, in two ways: (1) the accretion rate on the galaxy is reduced \citep[][]{Mitchell2018,Tollet2019}, and the covering factor of inflowing gas is decreased. This phase of complete blowout is typical of low-mass galaxies in which the energy injected by supernovae can easily balance the relatively weak gravitational binding energy of the system. The explosive sequence of panels (2) and (3) is accompanied by a disruption of star-forming clouds, which is not visible here. Once the galaxy has ejected most of its gas and delayed further accretion, its star formation rate slows down, and the explosions of supernovae cease to manage to generate a galaxy-scale wind. This allows CGM gas to cool and fall back, and we see in panel (4) that accretion streams have reformed, roughly 200 Myr later. At this point, the galaxy does not produce very red line profiles anymore, and instead it can be seen with very blue lines in $\sim30\%$ of the directions of observation. This fraction further rises to $\sim 40\%$ during the next 100 Myr, where the accretion flows further develop and star formation is not sufficient yet to produce a new galactic wind. After $\sim 200$ Myr of accumulating gas in the ISM, star formation becomes intense enough to trigger a new total blowout of the galaxy, and panels (5) and (6) show this phase which is very similar to that seen in panels (2) and (3) before. This completes the cycle of accretion and ejection phases and shows that indeed there is a strong correlation between these phases and the production of extreme red/blue line profiles. 

An important feature to note here is that the blowout is extreme in the sense that it pushes the gas at all scales and in most directions. This is probably much more significant in our low-mass galaxy than it would be for more massive galaxies. Note that the second peak of star formation is different from the first in the sense that star formation is not shut off as much because of feedback. We thus see a hint of what may happen in more massive galaxies: a cohabitation of the two inflowing and outflowing phases, with thinner, pressure-confined accretion streams, and volume filling, continuous outflows. In these more massive galaxies, a larger number of sites of star formation take turn to sustain both intense star formation and galactic winds, and we expect they will produce mostly very red and ordinary line profiles. Of course, more massive galaxies will also contain more dust, and this may change the picture to some extent. Our speculations here need to be tested in future work.

\subsection{The emergence of the \lya{} line profile}\label{sec:cgm}

We have seen that the presence of very red or very blue line profiles is connected to times and directions where either outflows or inflows dominate at all scales, from within the ISM to beyond the virial radius. We now investigate in more details the physical origin of these trends.

We start by defining a virtual boundary at $0.2\times R_{\rm vir}$ $(\sim 5$~kpc) from the DM halo centre to separate the galaxy and its CGM. This radius is roughly twice the radius containing 90\% of the stellar mass at any time, so that the full ISM is within this sphere, and most of the CGM is outside. Note that there is some CGM material within $0.2\times R_{\rm vir}$, in particular above and below the disc. We will thus refer to this region as the ISM and inner CGM in what follows, or loosely as the galaxy. In order to understand how the \lya{} line forms, we have repeated our RASCAS runs with the same observing strategy as before, but stopped the computation at $0.2\times R_{\rm vir}$. These runs provide us with the line profile emerging from the ISM and inner CGM, which we discuss in Sec. \ref{sec:ismspec}. Then, comparing these to our original mocks at $3\times R_{\rm vir}$, we discuss in Sec. \ref{sec:cgmspec} the impact of the CGM on the observed \lya{} profiles.

\begin{figure*}
\includegraphics[width=\textwidth]{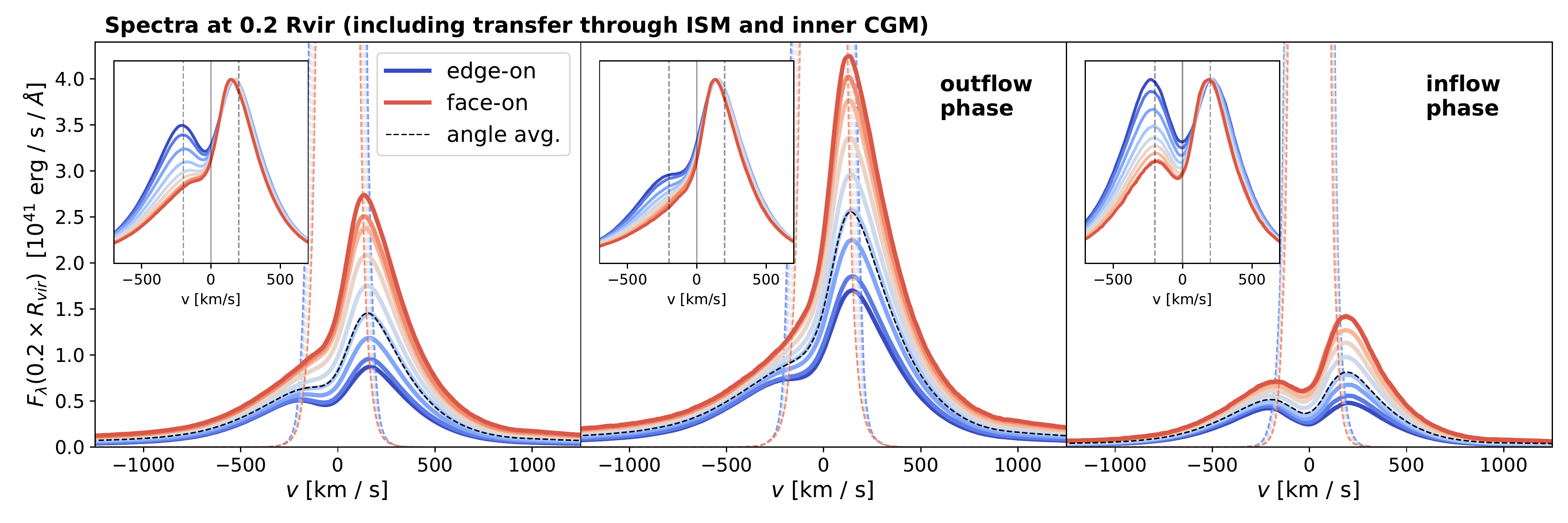}
\caption{The thick solid lines show the mock spectra observed at $0.2\times R_{\rm vir}$, stacked in regular bins of angle between the direction of observation and the angular momentum of the gas within $0.2\times R_{\rm vir}$. The colours show this angle, in steps of $\sim 11$ degrees, with blue lines indicating edge-on mocks and red lines face-on mocks.The left panel is a stack including all output times of the simulation, the central and right panels show stacks of mock spectra taken during outflow and inflow phases respectively (see text). In each panel, the thin dashed coloured lines show the stacked intrinsic spectra, and the black dashed line shows the angle average spectrum. The inset in each panel shows the same spectra, normalised to a their maximum value so as to make more apparent the line shape changes. The vertical lines in these insets highlight velocities of $-200$~km/s, $0$~km/s, and $200$~km/s to guide the eye. }
\label{fig:ISMSpectra} 
\end{figure*}

\subsubsection{The \lya{} line emerging from the galaxy} \label{sec:ismspec}

In Fig. \ref{fig:ISMSpectra}, we show the mock spectra obtained at $0.2\times R_{\rm vir}$, stacked in regular bins of inclination angle\footnote{As before, the inclination is measured relative to the angular momentum of gas within $0.2\times R_{\rm vir}$.}. Let us first focus on the left panel, which shows stacks of the mocks produced at all the output times of the simulation. 
We see that the escaping \lya{} luminosity varies with inclination so that face-on directions are generally brighter than edge-on views. Such a behaviour was already noted in previous works \citep[e.g.][]{Laursen2009,Yajima2012}, and discussed in detail in the context of idealised disc galaxies \citep{Verhamme2012,Behrens2014,Smith2022b}. In particular, \citet{Verhamme2012} showed that the variation of \lya{} luminosity with inclination is due to resonant scattering which leads \lya{} photons to escape towards the path of least resistance, i.e. rather perpendicular to the disc. We also find a correlation between line shape and inclination, such that edge-on mocks have more pronounced blue peaks than face-on mocks (see the inset in the left panel of Fig.~\ref{fig:ISMSpectra}). These results are again in line with those of \citet[][their Fig. 5]{Verhamme2012}. This is remarkable, because the simulation we use here differs in many important ways from the one they used. Our simulation evolves a morphologically complex galaxy in the cosmological context, while they simulated an idealised disc galaxy. Our simulation includes a full treatment of radiation-hydrodynamics and predicts the ionisation state of the gas and its \lya{} emissivity. Their simulation used hydrodynamics, assumed collisional ionisation equilibrium in post-processing, and emitted \lya{} photon packets from young star particles. The sub-grid models for star-formation and SN feedback used here are significantly different than those used by \citet{Verhamme2012}. They analysed sight-lines from a single output and here we stack 22,500 mocks that cover a full sequence of star formation and outflow. This suggests that the physical origin of the trend seen in Fig. \ref{fig:ISMSpectra} is quite robust and does not depend much on the detailed morphology of the galaxy, its evolutionary stage, or on the numerical methods used in the simulation. Our results thus confirm the findings of \citet{Verhamme2012} and extend them to a much more general context.
It also appears in Fig. \ref{fig:ISMSpectra} that the \lya{} lines emerging from the galaxy are generally red and quite broad. In particular, we find that the lines formed at $0.2\times R_{\rm vir}$ have a red peak that is shifted to $\sim 150$~km/s, very similar to what was found in \citet{Garel2021} when stacking \lya{} spectra of {\sc Sphinx} galaxies at $z>6$.

\begin{figure}
\includegraphics[width=\columnwidth]{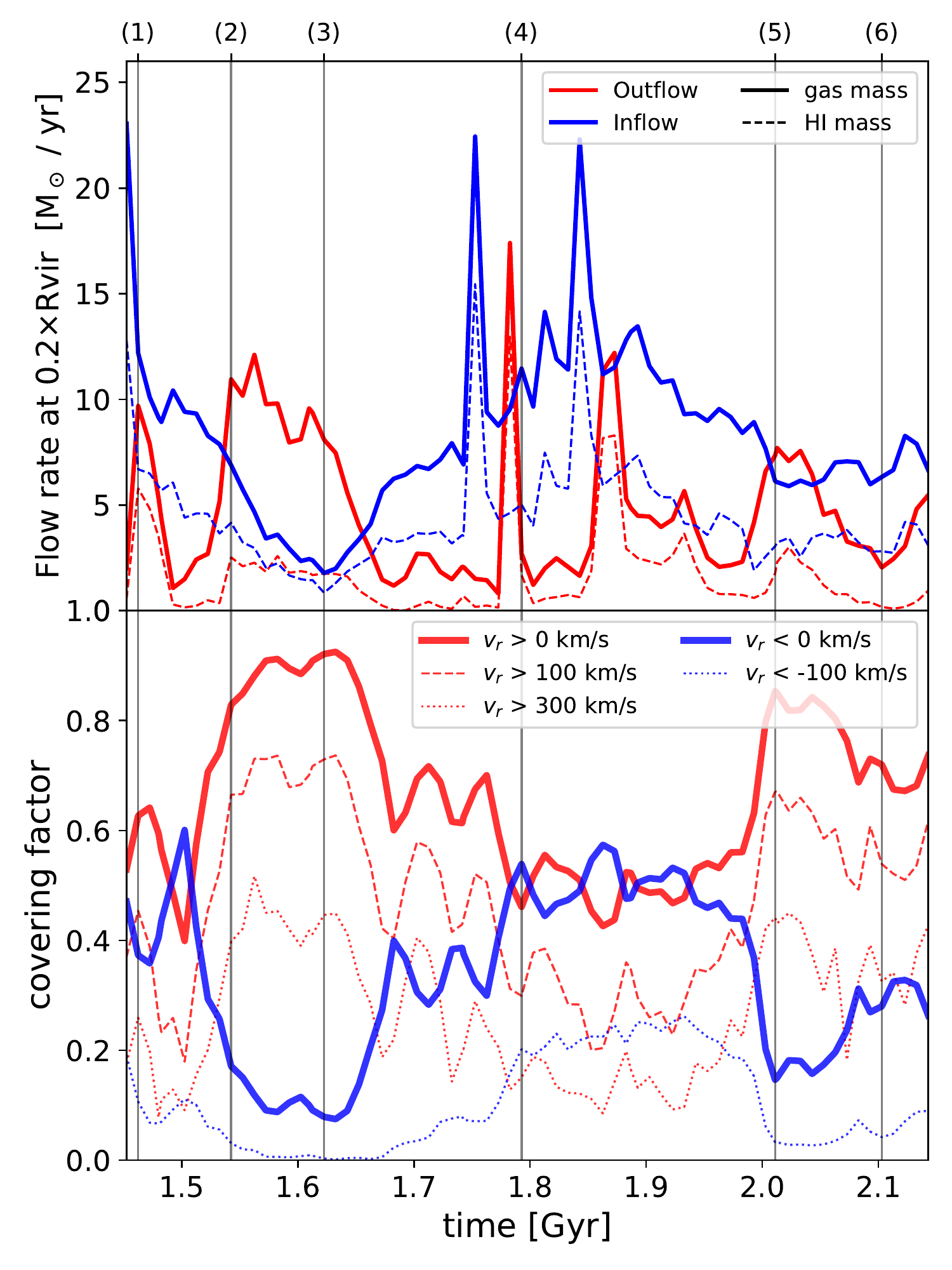}
\caption{{\it Top panel:} rates of mass flowing through a sphere of radius $0.2\times R_{\rm vir}$. The thick solid (thin dashed) lines show the total (HI) mass rates, with positive radial velocity in red, and negative radial velocities in blue. The two narrow peaks in the inflow rate at times $\sim 1.75$ Gyr and $\sim 1.85$ Gyr, followed by peaks in outflow rates shortly after, correspond to the passage of dense gas clumps through the $0.2\times R_{\rm vir}$ sphere before they are disrupted and fall back.  
{\it Bottom panel:} covering factor of outflows (red curves) and inflows (blue curves). The thin lines show the covering factors of high velocity components as indicated in the legend. The vertical lines mark the times illustrated in Fig. \ref{fig:sequence} for reference.}
\label{fig:ISMFlows} 
\end{figure}

In order to understand further these line shapes, we show in Fig.~\ref{fig:ISMFlows} the nature of the flows of gas through the $0.2\times R_{\rm vir}$ sphere as a function of time. Both the mass rates and covering factors of inflowing and outflowing gas are in line with the discussion of Sec.~\ref{sec:timesequence}, even though the measurement is made here at $0.2\times R_{\rm vir}$ only. In the lower panel of Fig.~\ref{fig:ISMFlows} we see that outflowing material always covers a significant part of the sky -- typically more than half -- even when the flows are dominated in mass by accretion. Together with the fact that outflow phases generally correspond to brighter emission (see below), this explains why the stacked line profiles on the left panel of Fig.~\ref{fig:ISMSpectra} are always red. 
The gas dynamics measured at $0.2\times R_{\rm vir}$ correlate very well with the line fractions of Fig.~\ref{fig:LineTypes_vs_time}. In particular, the times when the fraction of very red profiles is high in Fig.~\ref{fig:LineTypes_vs_time} correspond to strong outflow phases, both in terms of mass rate and covering factor of the outflowing material. 

In the centre panel of Fig.~\ref{fig:ISMSpectra}, we show stacked mock spectra emerging from the galaxy at times when more gas flows out of the $0.2\times R_{\rm vir}$ sphere than in\footnote{We obtain a similar result when stacking outputs that have a covering factor of outflows larger than 70\%, regardless the mass flows.}. We see that these times produce brighter lines, mostly because they happen shortly after intense star formation events so that the intrinsic \lya{} emission is large (see. Fig.~\ref{fig:SFR_Mstar}). These lines are also redder -- their blue peaks are less pronounced -- as expected when resonant scattering occurs in an outflowing medium, and have a weaker variation with inclination, because the outflow is relatively isotropic. The effect is slightly amplified if we stack only the few outputs around time (3), shortly after 1.6~Gyr. At this time, the outflow is extreme in the sense that its covering factor is very high ($\sim 90\%$), and that it expels more H{\sc i} than the galaxy accretes. At that point, the lines emerging from the galaxy do not vary much with inclination at all and show only a moderate blue flux excess. 

In the right panel of Fig. \ref{fig:ISMSpectra}, we stack mock spectra of the galaxy selected at times when the inflow has a covering factor larger than 50\%. As expected, we obtain fainter and bluer lines here, including symmetric double peaks in the edge-on direction. The lower luminosities are again linked to the star formation history. The variation of the line shape with inclination is stronger here than for outflows (see insets), because the inflowing gas fills only part of the sky, mostly close to the disc plane, while almost half the directions are outflowing towards the poles. The relatively symmetric double peak profile obtained edge-on again reminds us of the results of \citet{Verhamme2012} and suggests scattering through a medium with relatively low radial velocity. We note that regardless of the direction, the red peak is found at $\approx 200$~km/s in the right panel, whereas it was located at $\approx 150$~km/s in the other ones, which suggests that \lya{} photons have to scatter through larger column densities during accretion phases than in general.

A remarkable point from the comparison of the centre and right panels of Fig.~\ref{fig:ISMSpectra} is that the lines produced face-on during inflow phases always have a larger blue-to-red flux ratio ($B/T$) than the lines produced edge-on during outflow phases. Thus while all evolutionary phases retain a monotonous variation of line shape and luminosity with inclination, the dynamics of the ISM and inner CGM define the accessible range of line shapes at any time, and the recent star formation defines the typical luminosity.

\subsubsection{The role of the CGM} \label{sec:cgmspec}

\begin{figure*}
	\includegraphics[width=\textwidth]{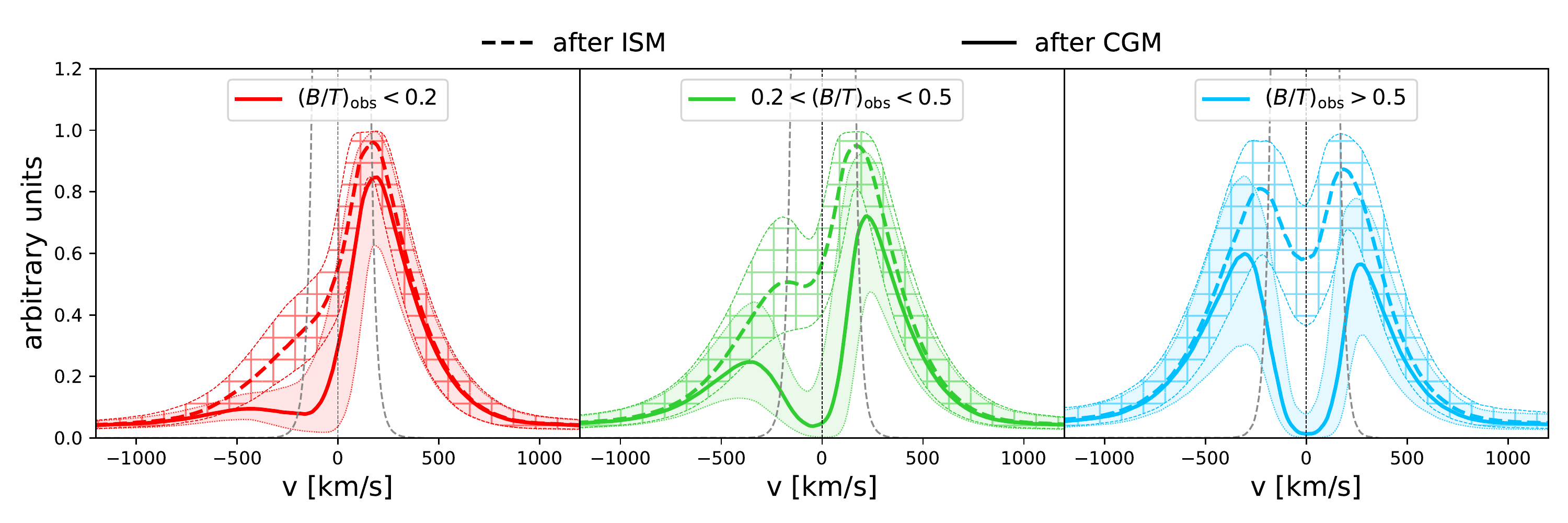}
    \caption{Median mock spectra emerging from the galaxy ($0.2\times R_{\rm vir}$, dashed lines) and from the CGM ($3\times R_{\rm vir}$, solid lines). The shaded or hatched envelopes are limited by the 15th and 85th percentiles. The different panels show the medians of spectra selected on their line profile as it emerges from the CGM. From left to right: very red lines ($B/T < 0.2$), ordinary double peaks ($0.2<B/T<0.5$), and very blue lines ($B/T>0.5$). In each panel, the grey dashed line shows the intrinsic \lya{} line, with the same normalisation as the other lines. }  
\label{fig:ISM_vs_CGM}
\end{figure*}

While the large-scale motions of gas in the CGM are strongly influenced by the galactic wind from the galaxy, we find that their complexity is such that the effect of the CGM, integrated from $0.2\times R_{\rm vir}$ to $3\times R_{\rm vir}$ does not correlate with the inclination angle of the galaxy. In order to understand the impact of the CGM in more detail, we thus approach the question from another angle: instead of stacking spectra in selected directions and times, we explore how the line-of-sight properties vary as a function of the observed spectral shape (at $3\times R_{\rm vir}$). In Fig.~\ref{fig:ISM_vs_CGM}, we show the median line profiles emerging from the galaxy and from the CGM for three sub-samples of mocks defined by the line colour $B/T$ after transfer through the CGM. We note the high level of consistency between the shapes of the lines emerging from the galaxy and from the CGM, which confirms that the observed line properties are rooted in the ISM and inner CGM as discussed in Sec.~\ref{sec:ismspec}. 

\begin{figure}
\includegraphics[width=\columnwidth]{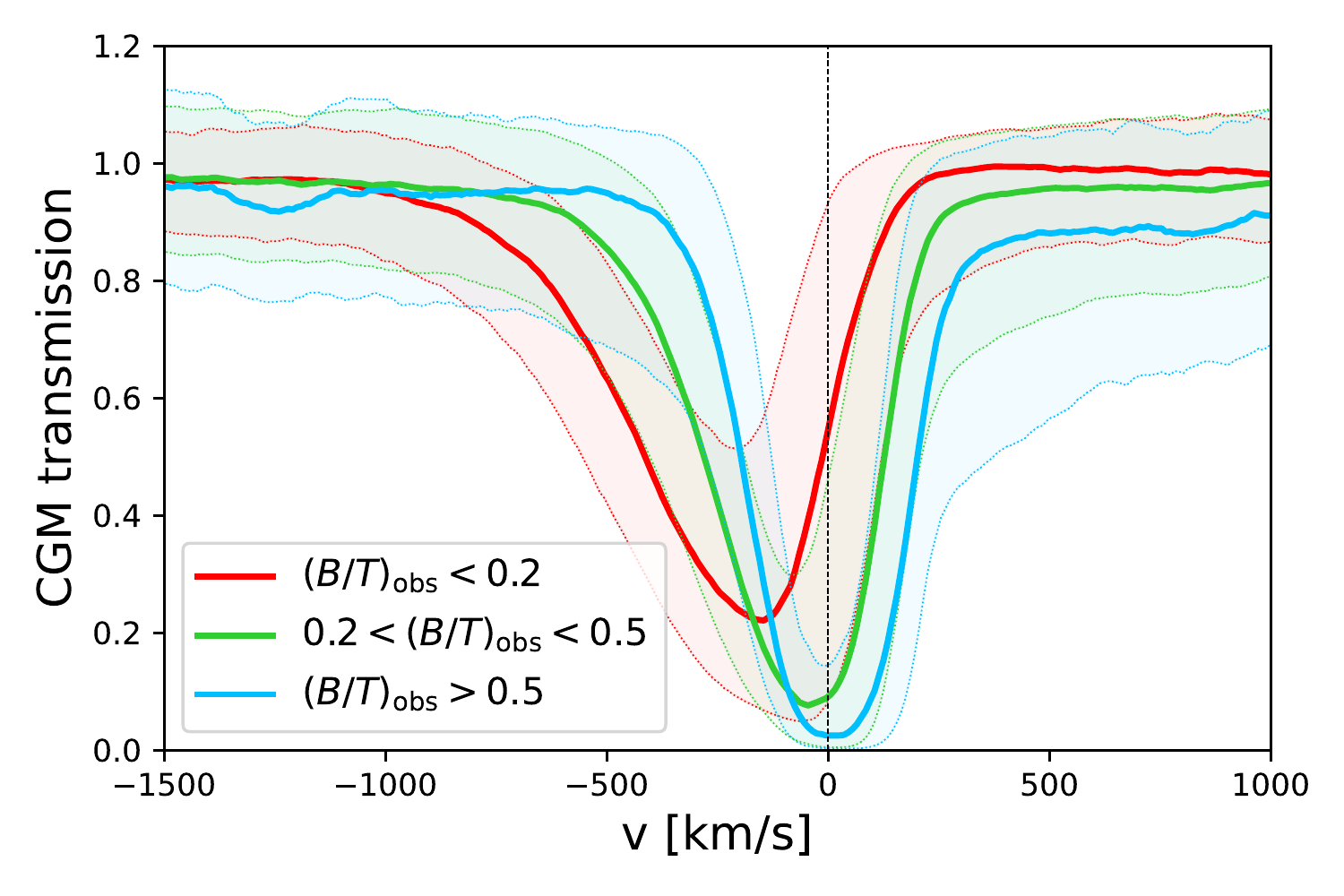}
\caption{Effective CGM transmission (see text) as a function of velocity for mock spectra grouped by their line colour $B/T$ as indicated on the plot. The thick lines show the median transmissions, and the boundaries of the shaded areas are the 15th and 85th percentiles.}
\label{fig:cgmtransmission} 
\end{figure}

Starting with the left-hand side panel of Fig.~\ref{fig:ISM_vs_CGM}, we see that lines which are observed with very red profiles emerge from the ISM as red-shifted and broad single peaks, typical of face-on views during outflowing phases (Fig.~\ref{fig:ISMSpectra}). Interestingly, the peak velocity does not change much between $0.2$ and $3\times R_{\rm vir}$, nor does the width of the line: the lines before and after CGM are indistinguishable in their wings. Here, the effect of the CGM is to mildly suppress flux in the blue side of the line, which is enough to carve a small valley and transform the single peak into a red-dominated, double peak profile, a P-Cygni profile, or a very asymetric single peak with a steep drop in flux on the blue side of the line. From the spectra $F_\lambda(r)$ emerging at different radii $r$, we can define an effective CGM transmission $\mathcal{T}_{\rm CGM}$ as the ratio between the flux emerging from the CGM to the flux entering the CGM: $\mathcal{T}_{\rm CGM} = F_\lambda(3\times R_{\rm vir}) / F_\lambda(0.2\times R_{\rm vir})$. In Fig.~\ref{fig:cgmtransmission}, we show the median transmissions measured for mock spectra with different line colours. The red curve shows that indeed the effect of the CGM on very red profiles is to produce a relatively shallow attenuation, peaking in the blue, and extending to very negative velocities (down to roughly $-1000$~km/s), as expected if the attenuation is produced by gas moving out at high velocity. 

Looking now at the central panel of Fig.~\ref{fig:ISM_vs_CGM}, we see that lines which are observed as ordinary double peaks emerge from the galaxy as either a single peak or a double peak with a very shallow central absorption trough. As seen in Fig.~\ref{fig:ISMSpectra}, such profiles are produced by the galaxy in all evolutionary phases and directions. Again, the width of the line is already in place when emerging from the galaxy, and the wings of the lines at $0.2\times R_{\rm vir}$ and $3\times R_{\rm vir}$ are superposed. The effect of the CGM here is again to produce an effective absorption line, peaking slightly in the blue, extending from $\sim -500$~km/s to $\sim 200$~km/s, and reaching saturation close to systemic velocity for a significant fraction of mocks. This CGM transmission strongly reduces the intensity of the blue peak emerging from the galaxy and shifts it further to the blue. It also reduces the red peak to a lesser extent, and shifts very slightly its velocity to the red.

The right-hand side panel of Fig.~\ref{fig:ISM_vs_CGM} shows that lines which are observed as very blue profiles generally escape the galaxy as a double peak with similar flux on the blue and red sides. From Fig.~\ref{fig:ISMSpectra}, we know these lines are produced by the galaxy in the edge-on plane and during major inflow events. At $0.2\times R_{\rm vir}$, the dip between the two peaks is not very pronounced yet, however, and the CGM does enhance this feature very significantly. Again, the width of the line is already in place out of the ISM and inner CGM, and the main effect of the CGM is to attenuate the flux at line centre. Fig.~\ref{fig:cgmtransmission} shows that the CGM transmission for very blue lines is a deep absorption line, often saturated, and peaking at slightly positive velocities. It is interesting that the effective transmission of the CGM seen in Fig.~\ref{fig:cgmtransmission} is largely responsible for the final separation of the blue and red peaks of the observed \lya{} line. This is particularly visible on the bluest lines (right panel of Fig. \ref{fig:ISM_vs_CGM}) where the broad absorption from the CGM at roughly systemic velocity increases the peak separation by $\sim 150$~km/s. In redder lines, because the effective absorption from the CGM is blue-shifted, the red peak velocity varies only a little while the blue peak is shifted to the blue by a few hundred km/s.  

\vskip 0.2cm 
In order to understand why the effective transmission of the CGM varies with sight lines, we compute for each mock observation the mass of gas and of H{\sc i} within a cylinder aligned with the line of sight, of radius 0.5 arcsec, and covering distances from $0.2\times R_{\rm vir}$ to $1.5\times R_{\rm vir}$ from the galaxy towards the observer. The section of the cylinder is the aperture in which we collect radiation to build the spectra. The range of distances we choose allows us to probe the CGM well outside the galaxy (of size $\sim 0.1\times R_{\rm vir}$), and to capture the large-scale motions of the CGM which extend beyond the virial radius (see Fig.~\ref{fig:sequence}). We further split this mass into three components depending on the line-of-sight velocity $v_{\rm los}$ of each gas element. We define $M_{\rm in}$ ($M_{\rm out}$) as the mass of gas in the cylinder inflowing with $v_{\rm los} < -v_{\rm thresh}$ (outflowing with $v_{\rm los} > v_{\rm thresh}$),  and $M_{\rm stat}$ the mass of gas with $|v_{\rm los}| \leq v_{\rm thresh}$ which we consider to be static. We use a fiducial value of $v_{\rm thresh} = 50$~km/s but also show some results obtained with $v_{\rm thresh}=90$~km/s which is close to the virial velocity of the halo. Along with mass, we compute for each sight-line the volume occupied by outflowing, inflowing, and static gas, and we define the relative volume fractions occupied by each component.  

\begin{figure}
\includegraphics[width=\columnwidth]{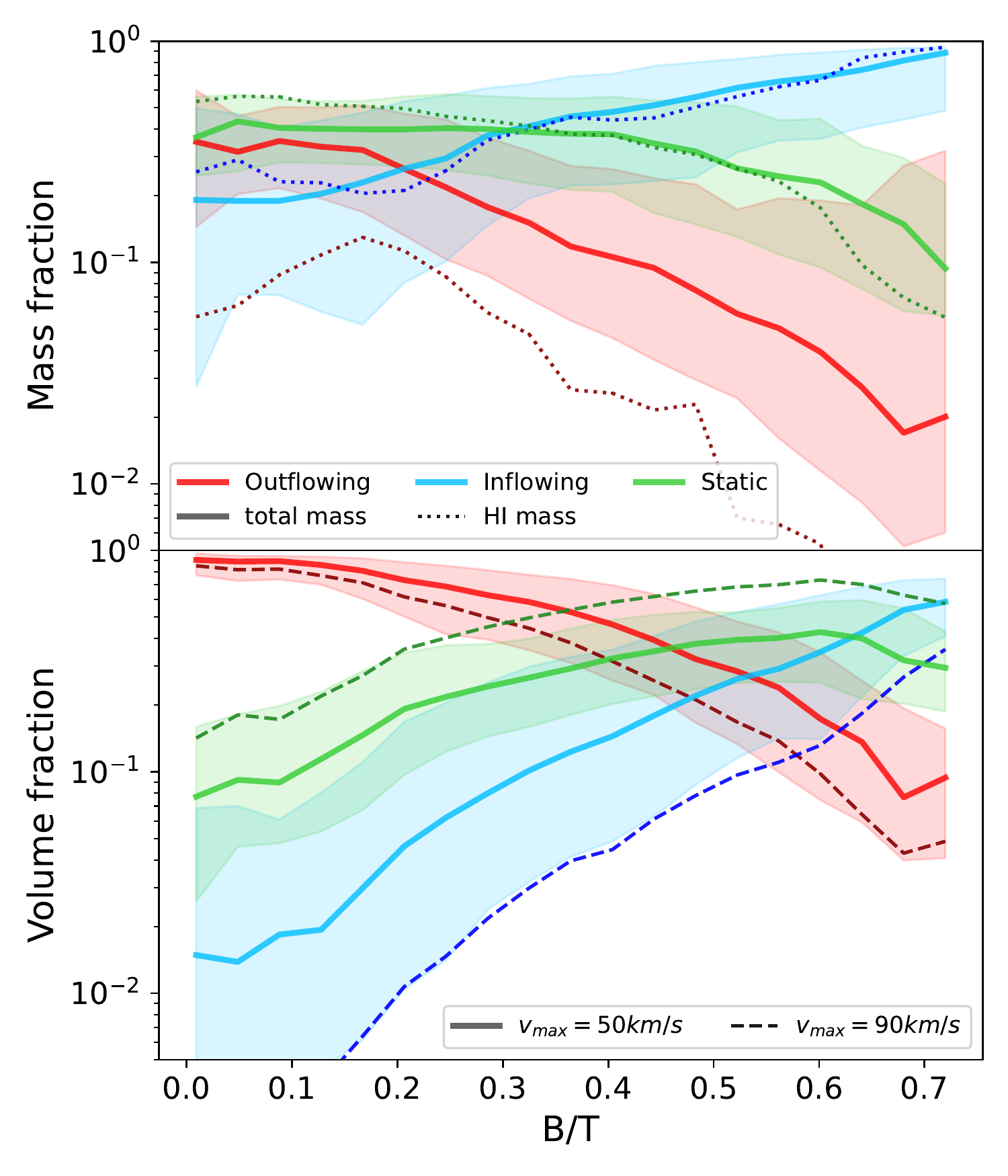}
\caption{{\it Top panel:} Mass fraction of gas in different flows on the line of sight, between $0.2\times R_{\rm vir}$ and $1.5\times R_{\rm vir}$, as a function of the blue-to-total flux ratio of the mock \lya{} lines. The points at $B/T=0$ include single-peak and P-Cygni profiles. The lines show the median values, and shaded areas cover the 15 to 85 percentiles. The solid lines show the median mass fractions, and the dotted lines the median H{\sc i} mass fractions. {\it Bottom panel:} Same as top panel but now with the volume fractions instead of the mass fractions. The dashed lines show the median values obtained with a higher velocity threshold to separate the flows ($v=90$~km/s). The solid lines (and dotted lines in the top panel) use the fiducial separation at $v=50$~km/s. In both panels, the red (resp. blue, green) lines show outflowing (resp. inflowing, static) gas. }
\label{fig:flows} 
\end{figure}

In Fig.~\ref{fig:flows}, we show how the masses and volumes of the three kinematic components are distributed as a function of the \lya{} line colour $B/T$. This figure shows a strong correlation between the colour of the \lya{} line and the relative mass and volume of each kinematic component of the CGM, with a qualitative behaviour in line with expectations from e.g. \citet{Dijkstra2006} or \citet{Verhamme2006}: blue profiles are associated to inflows, and red profiles to outflows. The quantitative details are worth discussing, however, and the picture which emerges from these results is more subtle than our first intuition. 

First, lines with a dominant blue peak ($B/T> 0.5$) are indeed found in sight-lines where the inflowing component dominates in mass (60\% to 90\%) and occupies a significant fraction of the volume (35\% to 50\%). These sight-lines also tend to contain very little outflowing gas: typically less than 7\% in mass, occupying a volume fraction lower than 30\%. This outflow component is mostly ionised and its column density and covering factor are thus very low. Indeed, the outflowing component is even more subdominant (lower than the per cent level) in terms of H{\sc i} budget, which suggests it has little impact on the \lya{} line. Conversely, the inflowing gas is largely neutral, and dominates the H{\sc i} budget more strongly. The inflowing gas does not move very fast: while it dominates the line-of-sight volume when we set $v_{\rm thresh}=50$~km/s, it is the static phase that becomes largely dominant when $v_{\rm thresh}=90$~km/s, showing that by volume, most of the inflow is at speeds lower than the virial velocity of the host DM halo. This behaviour is in line with the neutral and ionised gas flows measured in a similar simulation and discussed in much more detail by \citet{Mitchell2020}. From Fig.~\ref{fig:flows}, it thus seems that we can robustly associate lines with a dominant blue peak to situations where inflowing material largely dominates the line-of-sight CGM by mass and, to a slightly lesser degree, by volume. This dominant inflowing component is not very fast, which explains the CGM absorption peaking only slightly to the red side of systemic. 

Second, very red lines (including P-Cygni, single peaks, and double peaks with $B/T<0.2$) do {\it not} require a CGM dominated in mass by outflows. In fact, these lines are found in situations where  $\sim 40\%$ of the CGM is static, roughly 20\% is inflowing and only a bit more than 30\% is outflowing\footnote{These typical fractions are medians of the full distribution, and while the fractions per sight-line add up to unity, the medians need not do so.}. These fractions change significantly if we consider H{\sc i} only, instead of total mass, and the median outflowing H{\sc i} mass fraction reaches a maximum $\sim 15\%$ at $B/T\sim0.2$. Instead of mass, it is the volume fraction of outflows in the line of sight which seems to be the key driver here, and very red lines have in common a volume filling outflow, with more than 75\% of the volume occupied by material outflowing at $v_{\rm los}>50$~km/s. Along these sight-lines, the outflow is fast, and most of the volume is filled with gas leaving the galaxy at velocities larger than the escape velocity of the dark matter halo, i.e. $\gtrsim 90$~km/s. 

Third, the transition from very red lines to very blue lines happens as expected when the mass and volume fractions of inflows rise, and the mass and volume fractions of outflows decrease. However, the overall picture which emerges from Fig.~\ref{fig:flows} is more nuanced than most idealised models. The CGM generally appears to be a complex medium where very diverse kinematic components coexist. In particular, Fig. \ref{fig:flows} shows that regardless the line colour, a significant fraction ($\sim 40\%$) of the CGM mass is moving at low line-of-sight velocities. Most \lya{} lines in our sample are ordinary double peaks with 50\% to 80\% of the flux in the red peak, and they originate from sight-lines that have a fairly balanced mix of inflowing and static gas, but a very sub-dominant outflowing H{\sc i} component, at odds with common expectations from idealised models.

\vskip 0.3cm
From Figs.~\ref{fig:ISM_vs_CGM} and \ref{fig:cgmtransmission}, we see that scattering through the CGM does not broaden the \lya{} line at all. This tells us that most of the scattering processes responsible for significantly broadening the intrinsic emission line have already happened before the photons reach the CGM. Fig. \ref{fig:cgmtransmission} also shows that the transmission of the CGM reaches a value close to unity at low and high velocities, which means that there is no significant dust absorption in our simulated CGM. Indeed, the effect of dust is mostly concentrated at the ISM scales \citep[as in][]{Smith2022b} and is responsible for a mean, angle-averaged, \lya{} escape fraction of $\approx 8.2\%$ over time. The absorption feature in the transmission is thus due to photons being scattered off the line of sight rather than being absorbed by dust or redistributed in frequency. Because the aperture of 1~arcsec through which we collect photons to build mock spectra is much smaller than the full extent of the CGM, photons which scatter in the CGM are easily lost to the detector and do not contribute to the mock spectra. These photons would be recovered with sufficiently deep observations that integrate the \lya{} flux over larger apertures. Indeed, extended \lya{} emission is observed to be common around galaxies similar to the one we simulated \citep[e.g.][]{Wisotzki2016, Leclercq2017, Kusakabe2022}, and \citet{Mitchell2020} demonstrated that part of that signal comes from scattering of galactic photons in the CGM. The observations to which we compared our results in Sec. \ref{sec:LASDComparison} use an observational strategy similar to ours in the sense that the COS aperture is typically comparable to or smaller than the size of the observed galaxies. Part of our success in matching these observations does indeed come from using the same observational strategy and thus limiting the role of the CGM to producing an effective absorption line. Fig. \ref{fig:cgmtransmission} of course reminds us of the results of \citet{Laursen2011} who measured the effect of the intergalactic medium on the \lya{} line. These authors show a similar effect to what we find, and indeed the mechanism is similar. There are differences, however, that we discuss in Sec. \ref{sec:LaursenDiscussion}. 

\subsubsection{Summary} 

The picture which finally emerges is one in which the \lya{} line shape develops gradually from emission to observation, with the ISM and the CGM both leaving strong and different signatures. Scattering processes in the ISM and inner CGM (loosely defined as gas within $0.2\times R_{\rm vir}$) produce a broad line -- much broader than the intrinsic emission -- which is generally dominated by a red peak at $\sim 150$~km/s and may feature a smaller blue peak at $\sim -200$~km/s. The intensity and shape of the \lya{} line that emerges from the galaxy depends on the evolutionary phase of the galaxy: outflow phases that follow starbursts produce redder and brighter lines, and inflow phases produce weaker and bluer lines. The line shape also depends on inclination and at any time, the line colour varies monotonously from red to blue when observing the galaxy from face-on to edge-on directions, in a range of $B/T$ values set by the evolutionary stage. The line emerging from the galaxy is then processed by the CGM which acts as a screen that scatters some \lya{} photons out of the line of sight, producing a strong effective absorption feature. This broad absorption is more or less shifted to the blue depending mostly on how the volume-filling component of the CGM on the line of sight moves. It is this absorption by the CGM that primarily drives the separation of the blue and red peaks, while the global width of the line, as measured by its wings, is already set just outside the galaxy. The flows of gas that traverse the CGM along any sightline typically contain multiple phases and kinematic components. Very blue line profiles require a CGM which is dominated by relatively slow inflows. Very red lines require a CGM dominated in volume by fast outflows, but these typically represent a very sub-dominant part of the H{\sc i} budget. Ordinary sight-lines have mix of inflowing and outflowing gas, with most of the mass in the inflow but most of the volume occupied by the outflow. 

In this framework, the common interpretation of peak separation as being due to the scattering process highlighted by \citet{Neufeld1990} seems hazardous, and it is unclear how idealised models may constrain the properties of the CGM if they demand that scattering through a unique medium, be it multiphase, jointly produces the width of the line and its precise shape. Instead, our results suggest that because of the small aperture used to mock-observe the galaxy, the shape of the \lya{} line is due to the joint but different actions of the ISM and the CGM, so that the width of the line and depth of the absorption trough are not directly correlated.

\section{Discussion} \label{sec:discussion}
We have seen that the sample of mock \lya{} lines that we produce with our simulation contains most line shapes, and that their properties broadly cover the range of observed values. To our knowledge, it is the first time simulations reach this degree of realism, and this is in itself extremely encouraging. As we saw in Sec. \ref{sec:LASDComparison}, this introduces a change of paradigm in the way we can use simulations to interpret \lya{} observations, basically moving from a role of support for a qualitative interpretation to the possibility of quantitative interpretations by direct comparison to observations (see also Gazagnes et al., {\it submitted}, for a similar study focused on low-ionisation state metal absorption features). In this section, we wish to discuss further some elements of interpretation of our results. We start with a note of caution, then discuss the impact of selection effects, and finish with a discussion of earlier works.

\subsection{A note of caution} \label{sec:caution}

The agreement between our mock observations and real observations of low-redshift \lya{} emission lines is such that one is compelled to compare the distribution of mock properties to those which are observed, as we have suggested in Figs. \ref{fig:b2t_vs_vred}, \ref{fig:Orlinke}, and \ref{fig:vred_vs_fwhm}. This should only be done with caution, however, and the distributions of line properties from our sample of mocks cannot be directly compared to the distributions inferred from observations. There are two simple reasons for this which we reiterate below. 
 
First, we have investigated the properties of {\it a single galaxy}. This has allowed us to understand in some detail how \lya{} line properties are associated to different phases in the evolution of a galaxy and its CGM, or simply to the direction of observation relative to the gaseous disc of the ISM. It has shed light on some trends between line parameters that appear even when mock-observing a single galaxy, with roughly constant SFR and stellar mass. In order to understand how these trends translate to observations, one of course needs to study how they develop when looking at a representative sample of galaxies, which covers the full diversity of physical situations. It is clear, as we have discussed in Sec. \ref{sec:LASDComparison}, that our simulated galaxy is for example not a good match to the higher-mass Lyman break galaxies (or their low-redshift analogs). Investigating the \lya{} properties of a larger sample of galaxies will be the focus of future work using the {\sc Sphinx} cosmological simulations \citep{Rosdahl2022}. 
 
Second, we have not applied any selection to our mocks, e.g. on UV magnitude, line luminosity, or equivalent width. The distributions we have shown are thus not comparable to the observed ones, which always include such selections. Indeed, we find that luminosity or equivalent width cuts introduce strong biases, and we discuss this further in Sec. \ref{sec:ohlala} below. 

\subsection{Correlations between line shape, luminosity, and equivalent width} \label{sec:ohlala}

\begin{figure}
\includegraphics[width=\columnwidth]{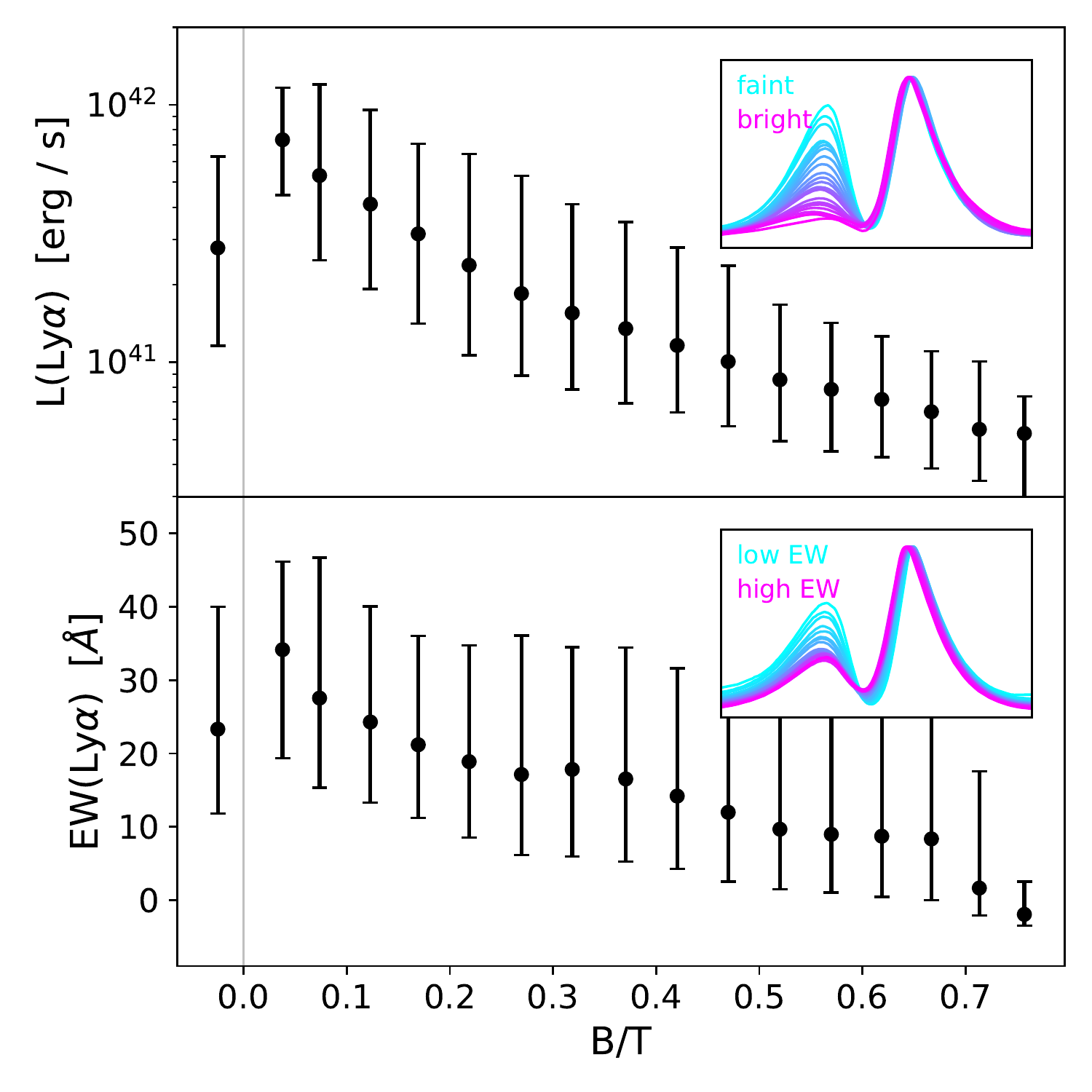}
\caption{Observed \lya{} luminosity ({\it top panel}) and equivalent width ({\it bottom panel} of the \lya{} line as a function of its colour $B/T$. In both panels, the points show the median values over our full sample of mocks, and the error bars indicate the 15th and 85th percentiles, in bins of $B/T$. The left-most point shows the values for other line profiles combined (single peaks, multiple peaks, and P-Cygni). The inset in each panel shows the mean normalised line profile for subsamples of mocks in bins of increasing luminosity (top panel) and equivalent width (bottom panel), from blue to purple. The samples are constructed so that each mean spectrum contains the same number of mocks. The $x$-axis of the insets run from $-1000$~km/s to $1000$~km/s.}
\label{fig:LumSelection}
\end{figure}

So far, we have chosen to discuss the shape of the \lya{} line without considering its luminosity or equivalent width (EW). However, because of the causal links between inflows, star formation, and outflows on the one hand, and because of the correlations that we have shown between the \lya{} line properties and these processes on the other hand, we expect that the line colour will be correlated with its apparent luminosity and equivalent width. In Fig. \ref{fig:LumSelection}, we show that indeed the luminosity and equivalent width of mock \lya{} lines are strongly correlated to their colour $B/T$: redder lines are on average brighter and have higher EW, while bluer lines are on average fainter and with lower EW. The values of luminosity and EW for other line types are a mixed bag and do not stand out in a noticeable way. Their medians are shown with the left-most points in Fig. \ref{fig:LumSelection}. Note that the definition of EW is always difficult for complex lines such as \lya{}, because the continuum around the emission line is generally not flat as \lya{} scattering produces an absorption feature that is much more extended than the emission line itself. We have chosen to measure the EW in our mocks as the excess of flux relative to the continuum integrated from $-1200$~km/s to $1200$~km/s. This velocity range is generally a good match to the line width and selects mostly the emission part of the signal. In order to estimate the continuum, we compute the median value of the flux at velocities with absolute values between $2000$~km/s and $5000$~km/s. 

The insets in each panel of Fig. \ref{fig:LumSelection} show how the spectral shape of the \lya{} line changes when computed for bins of increasing line luminosity and EW. This confirms that the trends hold also the other way: brighter (fainter) mocks are on average redder (bluer), and mocks with larger (smaller) EWs are generally redder (bluer). It is now important to remember that we are looking at a single simulated galaxy. While this galaxy goes through a few evolutionary phases, its stellar mass is roughly constant and its star formation rate too, if measured with UV broad-band photometry sensitive to variations on timescales of order $100$~Myr. The trends we see on Fig. \ref{fig:LumSelection} may thus hold for a population of fixed mass and SFR, but it is unclear what they would yield for a representative population of galaxies. Confirming this result observationally will be challenging and demands very deep observations that are complete in \lya{} for a mass-selected sample. On the theoretical side, expanding this result to a representative population of galaxies is no less challenging, although the {\sc Sphinx} simulations \citep{Rosdahl2022} offer a promising route.

Fig. \ref{fig:LumSelection} has other very strong observational implications. It suggests that \lya{}-selected galaxies will preferentially consist of galaxies undergoing strong outflows after strong episodes of star formation. This may bias our interpretation of what happens in the Universe and lead us to overestimate the prevalence of outflows. The trends in Fig. \ref{fig:LumSelection} may also explain in part the rarity of blue-peak dominated profiles in observations. Interestingly, our results also imply that stacks of UV-selected samples may  strongly favour red profiles, because they are brighter. Stacking UV-selected galaxies may thus not be a certain means to obtain line profiles which are representative of the full population of galaxies with a given SFR. We note however that a UV selection will also be affected by dust attenuation, so that UV-bright galaxies (at fixed SFR) may also be \lya{}-bright. Further analysis is needed to understand fully how UV and \lya{} selections interact to bias our vision of distant galaxies.

\subsection{Discussion of earlier works} \label{sec:LaursenDiscussion}
We have pointed out that previous attempts to simulate the \lya{} line, including our own, did not achieve as realistic \lya{} spectra as the ones shown in the current paper. We now identify which specific points of our work have made the difference.  

Very much like the mechanism underlined by \citet{Laursen2011}, the origin of the shape of the \lya{} line in our mock observations can be attributed coarsely to two elements. First, radiative transfer through the ISM broadens significantly the intrinsic emission line, and generally produces a red peak at $\sim 150$~km/s, with possibly a blue peak of lower intensity at roughly opposite velocity. The line which emerges from the ISM already has its final width, but its shape is not like observed lines and its absorption trough is very shallow when present. It is then the CGM that comes in to produce the final shape, by imposing an effective absorption line onto the spectrum emerging from the ISM. The impact of this absorption ranges from carving a deep absorption trough close to zero velocity, to producing a relatively shallow attenuation extended far out in the blue. Generally, the effect of the CGM is thus to enhance the asymmetry of the red peak, to redden the line by suppressing flux on the blue side, and to generate well-separated double peaks by producing a strong absorption trough close to systemic. 

While our analysis is thus in line with the general picture laid out by \citet{Laursen2011}, our conclusions differ in numerous ways. One important difference is that we do find that the typical spectrum emerging from the ISM is generally red, with little flux on the blue side. This is consistent with the results of \citet{Verhamme2012}, but somewhat in disagreement with \citet{Laursen2011} who always find an important blue component just outside their ISM (see their Fig. 7 for intermediate and massive galaxies). Hence, part of the success of our mock spectra comes from our ISM model, and to the strong SN feedback model we use here in particular. Another important point is that in our simulation, the galactic winds dominate over cosmological expansion far outside of the virialised dark matter halo, and indeed, we have obtained our results while neglecting the Hubble flow. The fact that the CGM contributes significantly to the \lya{} profiles may feel uncomfortable, because we have used a galaxy simulated at $z=3-4$ to represent observations at $z\sim0$. If the densities of outflowing and inflowing gas were set by cosmology and varied as $(1+z)^3$, we would indeed expect a significant difference in the effect of the CGM between redshifts 0 and 3-4. For the relatively low-mass galaxy that we simulate here, however, we find that the dynamics of the outflowing gas overwhelm cosmological expansion, so that the density of the expanding medium does not vary with redshift as naively expected: the galactic wind is responsible for the blue-shifted absorption, not the expanding IGM. The scenario we have outlined thus likely holds for small starbursting galaxies at low redshifts, even when the IGM is transparent. The physiognomy of inflows in low-mass dark-matter halos may change from high to low redshift, but we expect that this will change the occurrence rate of redshifted absorption rather than the overall mechanism that we have described. Thus, while the distribution of line shapes from our simulation is again not comparable to the observed one at low redshifts, the mechanisms that determine this distribution seem robust, and a key to the success of our mock spectra is that they account for \lya{} RT in a well resolved ISM {\it and} through the extended CGM. 

\vskip 0.2cm
Another important factor in our work is undoubtedly the relatively small aperture within which we have collected radiation to build mock spectra. This choice is deliberate, and meant to reproduce coarsely what is done by most low-redshift observations where the aperture of the COS spectrograph, for example, is often comparable to, or smaller than, the target galaxies \citep[see e.g.][for the recent CLASSY survey]{Berg2022}. Collecting light in a small aperture, compared to the $3\times R_{\rm vir}$ sphere within which we compute the \lya{} RT, favours the simple absorption patterns from the CGM that we have seen, because most of the scattered light does not come back on the line of sight and into the detector. This scattered light may be seen in observations carried out with integral field spectrographs such as MUSE or KCWI \citep[e.g.][]{Leclercq2020,Erb2022}. We have shown in \citet{Mitchell2020} that extended \lya{} emission is indeed powered partly by scattering from a central source, but also by in-situ emission and emission from faint satellites. A full analysis of the spectra of these different contributions deserves a future paper.

\vskip 0.2cm 
Finally, we note that our simulation of course has a limited resolution, which is much coarser than the one achieved in the recent idealised experiments of \citet{Kakiichi2021} or in the molecular cloud simulations of \citet{Kimm2022}. Interestingly, we find that the lines emerging from our simulated galaxy (at $0.2\times R_{\rm vir}$) are generally broader than reported in these works. Producing large velocity shifts is not hard in principle, if \lya{} photons have to scatter their way out of the ISM to escape. For example, Eq. (2) of \citet{Kimm2022} shows that with only moderate H{\sc i} column densities $N_{\rm HI}\sim 10^{19}$~cm$^{-2}$ and slightly supersonic turbulence (e.g. with Mach number $\mathcal{M}\sim 2$) scattering will already produce a red peak at $v_{\rm red}\sim 200$~km/s, larger than what we find in general \citep[see also the discussion of][]{Kakiichi2021}. Our results, in comparison to these higher-resolution studies, suggest that the diffuse ISM and inner CGM, which they cannot account for, many not have a negligible role in terms of scattering. 

Because the line shape relates to the dynamics of the ISM and inner CGM as well as to the more extended CGM, we expect that feedback models will have an impact on the colour of the lines emerging from the galaxy. In particular, models including non thermal pressure from cosmic rays have been shown to produce colder outflows \citep[e.g.][]{Farcy2022}, which could lead to redder lines in the face-on direction. Our comparison with observations in Sec. \ref{sec:LASDComparison} does not indicate that we require such feedback models, however, and one will need to compare the statistical properties of the \lya{} line for large samples of galaxies -- simulated and observed -- in order to constrain feedback models better.

\section{Conclusions} \label{sec:conclusions}

In this paper, we have produced mock \lya{} observations using a simulated high-redshift galaxy, typical of low-mass Lyman-$\alpha$ Emitters (LAEs). We have obtained the following results.
\begin{itemize}
\item Our mock \lya{} observations from a single simulated galaxy are able to reproduce the detailed shape of most of the \lya{} lines observed in galaxies of the local Universe, with a level of accuracy comparable to popular idealised shell models (Sec. \ref{sec:LASDComparison}). The observed lines that our mocks do not produce either correspond to galaxies much more massive than our simulation, or to rare and extreme LyC leakers with very peculiar properties. 
\item The line shape depends on the direction of observation relative to the galactic disc. Lines with more than 80\% of the flux in the red peak are preferentially observed face-on, while lines with a dominant blue peak are preferentially seen edge-on. However, most lines have intermediate values of the blue-to-total flux ratio ($0.2<B/T<0.5$), and are seen in all directions with no preference. 
\item The line shape also varies with time, following important evolutionary phases of the galaxy. Lines with a dominant blue peak ($B/T>0.5$) appear when accretion flows dominate. Very red profiles ($B/T<0.2$) are produced when outflows dominate the volume on the line of sight. For the low-mass galaxy that we have simulated, these phases come in turn so that blue and red profiles are not seen at the same times. Ordinary double peak profiles ($0.2<B/T<0.5$), however, are again present at all times in a roughly constant fraction of directions, and do not inform us on the evolutionary phase of the galaxy.
\end{itemize}

We have further discussed the physical origin of these trends and shown that the variations of the \lya{} line shape and luminosity with time and direction are generated by different effects through the galaxy and its CGM. In particular, we have found the following results. 
\begin{itemize}
\item The line emerging from the galaxy (at $0.2\times R_{\rm vir}$) is already broad and generally red, with a strong red peak at $\approx 150$~km/s, and a weaker blue peak at $\approx -200$~km/s.
\item At any time, the luminosity and shape of the line emerging from the galaxy vary monotonously with inclination: mocks are brighter face-on than edge-on, and the blue-to-total flux ratio $B/T$ is lower face-on than edge-on.
\item The luminosity and shape of the \lya{} line emerging from the galaxy also vary in time, in a way which is strongly correlated to the flows of gas traversing the $0.2\times R_{\rm vir}$ sphere. During outflow phases, the line is bright and red (low $B/T$ values), while during inflow phases the line is faint and blue (high $B/T$ values). The luminosity variations are mostly driven by the star formation history and the line shape by resonant scattering within $0.2\times R_{\rm vir}$. 
\item The line emerging from the galaxy then propagates through the CGM, which acts as a screen that produces a broad absorption line. This behaviour is due to the fact that we construct our mock spectra using a relatively small aperture of diameter 1~arcsecond, which is not much larger than the stellar body of the simulated galaxy. Thus, photons that scatter in the CGM are generally lost and do not contribute to the mock spectra.  
\item The absorption line produced by the CGM varies with the kinematic properties of the gas and strongly affects line properties such as peak separation or residual flux. Sight-lines where outflows dominate in volume produce broad blue-shifted absorption lines that attenuate the \lya{} lines emerging from the galaxy mostly on the blue side. This results in very red profiles with a red peak velocity set by radiative transfer through the galaxy and a weak blue peak far in the blue. Sight-lines which are dominated by inflowing material produce deep absorption lines close to the systemic velocity. These attenuate the lines produced by the galaxy around zero velocity and, by doing so, push the blue and red peaks further out and increases the peak separation by $\sim 150$~km/s. 
\item While our results follow the general expectation that outflows correspond to red line profiles and inflows to blue ones, our analysis shows a more nuanced picture. Very red \lya{} lines are found in directions where most of the CGM volume is occupied by outflowing gas. However, most of the mass in these lines of sight is generally inflowing or has no significant radial motion. Conversely, very blue lines are found in sight-lines in which the inflowing component dominates both the mass and volume budget. \end{itemize}

We have also briefly discussed some observational implications of our results, showing that our mocks predict a strong correlation between \lya{} luminosity and line colour, or between the equivalent width of the \lya{} line and its colour. These correlations are natural outcomes of the results above and suggest potentially strong biases. In particular, \lya{}-selected samples of galaxies may overestimate the prevalence of outflows in high-redshift galaxies.  

While the results presented here are very encouraging, further work is needed before we can use them to interpret more quantitatively deep \lya{} surveys. The series of {\sc Sphinx} simulations \citep{Rosdahl2022} is the ideal tool to give a statistical perspective to our findings and explore their implications for the observation of distant galaxy populations.

\section*{Acknowledgements}
We thank the anonymous referee for constructive comments that helped improve this paper. We gratefully acknowledge support from the PSMN (Pôle Scientifique de Modélisation Numérique) of the ENS de Lyon for the computing resources. The simulation used in this work and the \lya{} RT computations were carried out on the Common Computing Facility funded by the LABEX Lyon Institute of Origins (LIO, ANR-10-LABX-0066) and hosted at the PSMN. 
AV is supported by the SNF Professorships PP00P2\_176808 and PP00P2\_211023. AV and TG are supported by the ERC Starting Grant 757258 "TRIPLE".
TK was supported by the National Research Foundation of Korea (NRF) grant funded by the Korea government (No. 2020R1C1C1007079 and No. 2022R1A6A1A03053472).
MT acknowledges support from the NWO grant 016.VIDI.189.162 ("ODIN").
This work was supported by the Programme National Cosmology et Galaxies (PNCG) of INSU, CNRS with INP and IN2P3, co-funded by CEA and CNES.

\section*{Data Availability}
The mock spectra are available upon request. 


\bibliographystyle{mnras}
\bibliography{library}




\appendix

\section{Matches to the LASD}\label{app:allfits}
\begin{figure*}
	\includegraphics[width=\textwidth]{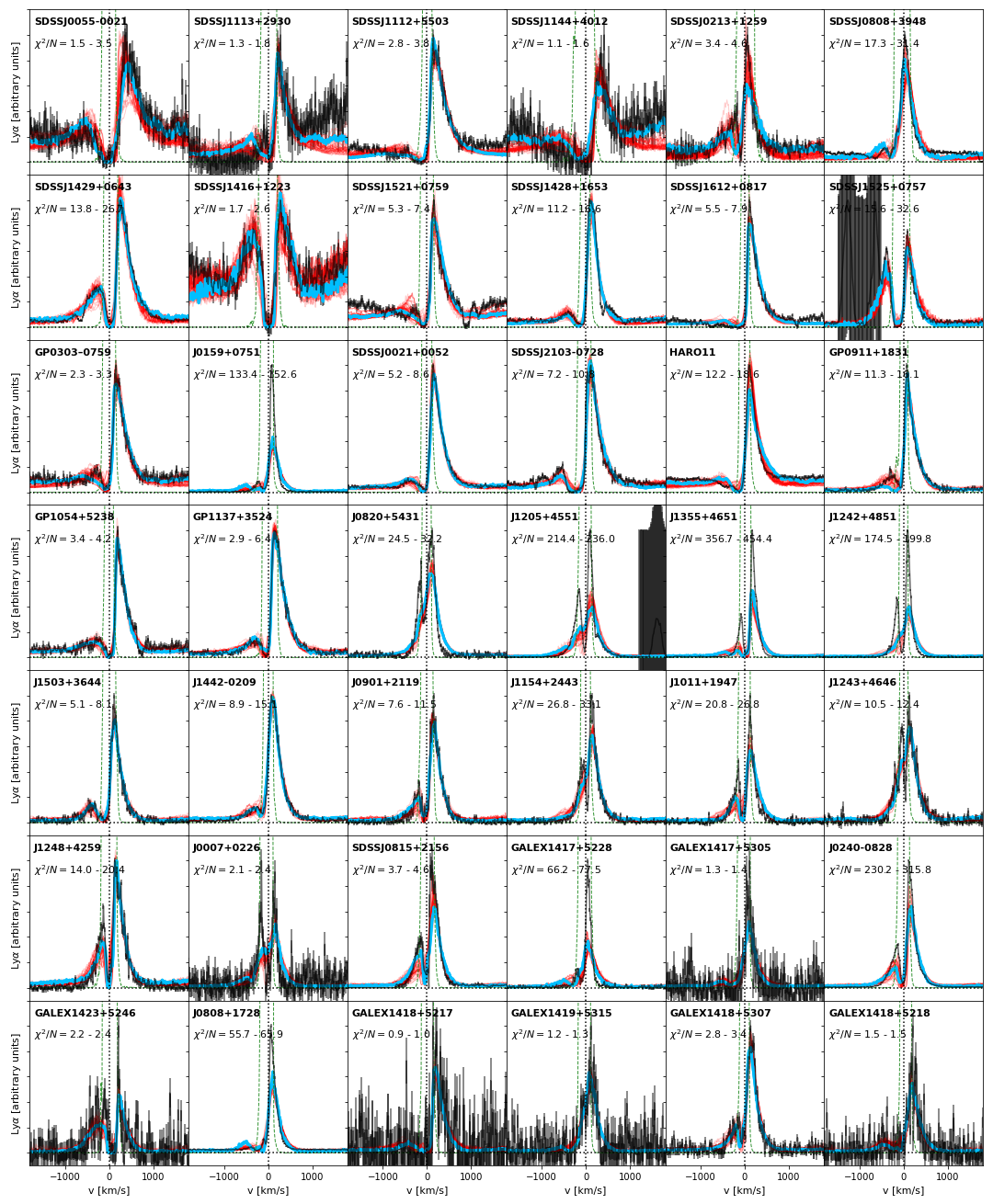}
    \caption{Same as Fig. \ref{fig:SuccessfulFits}, but for other spectra from the LASD.}  
\label{fig:allfitsfirst}
\end{figure*}

\begin{figure*}
	\includegraphics[width=\textwidth]{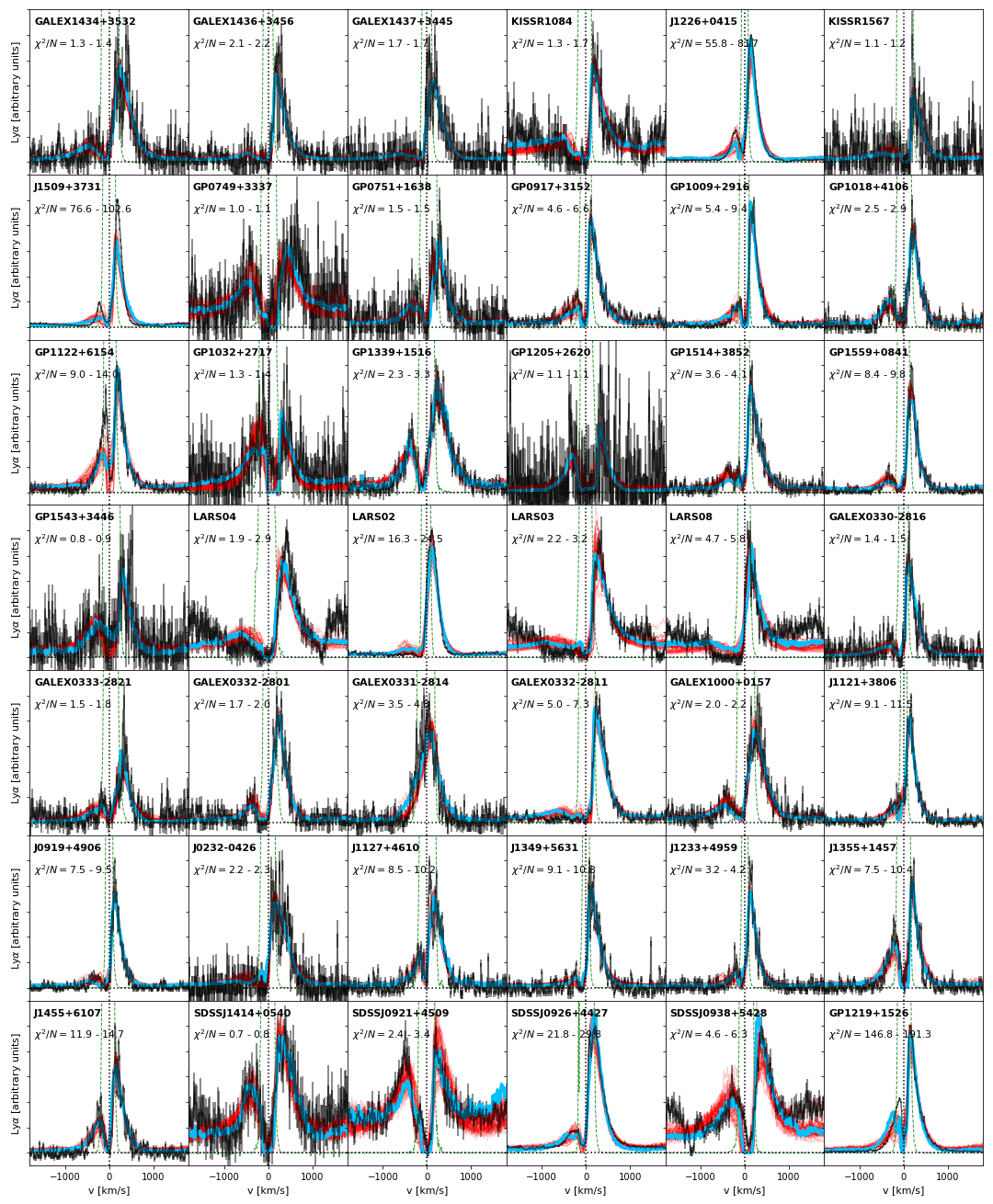}
    \caption{Same as Fig. \ref{fig:allfitsfirst}.}  
\label{fig:allfitssecond}
\end{figure*}

\begin{figure*}
	\includegraphics[width=\textwidth]{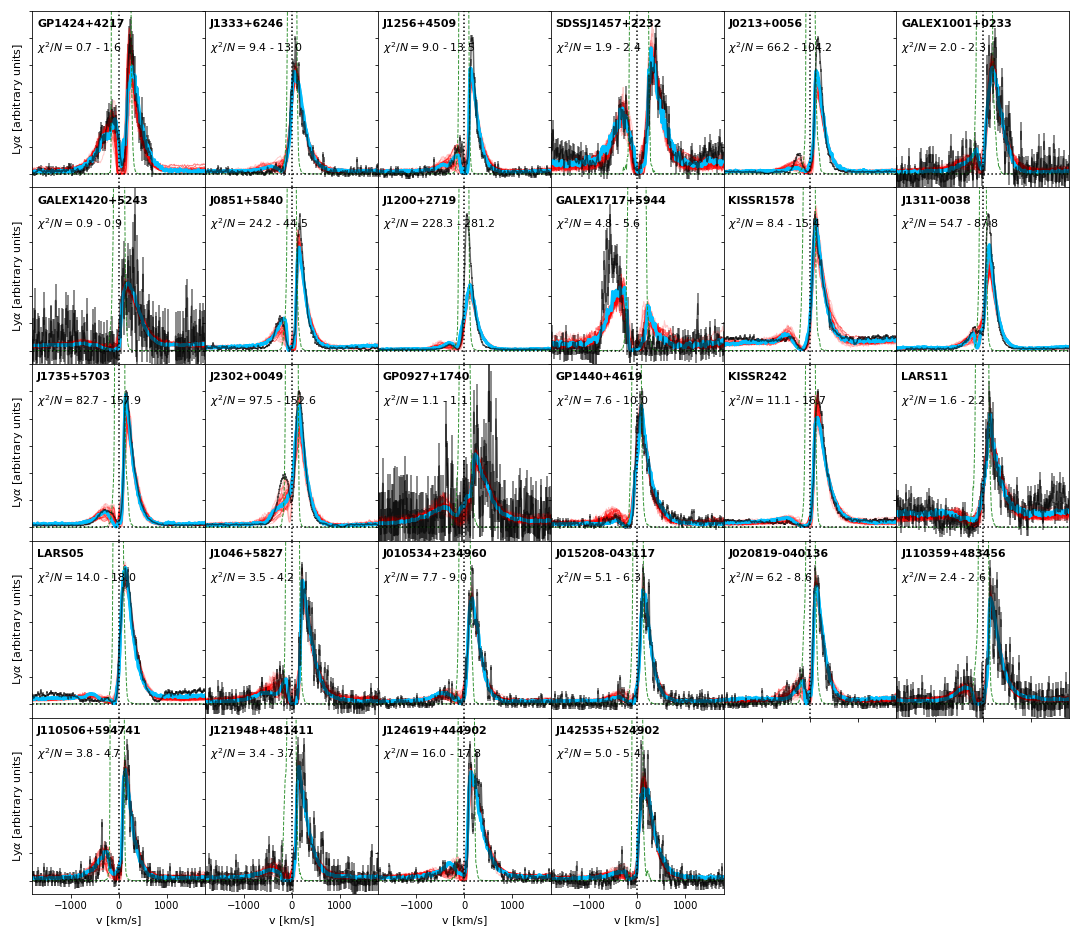}
    \caption{Same as Fig. \ref{fig:allfitsfirst}.}  
\label{fig:allfitsthird}
\end{figure*}

In Figs. \ref{fig:allfitsfirst} to \ref{fig:allfitsthird}, we show our best-matched mocks to all low-$z$ spectra from the LASD, except for those already shown in Figs. \ref{fig:SuccessfulFits} and \ref{fig:FailedFits}. 

\bsp	
\label{lastpage}
\end{document}